\documentclass[a4paper,11pt]{article}
\usepackage{jheppub} % for details on the use of the package, please see the JINST-author-manual
\usepackage{lineno}
\usepackage{amssymb,amsfonts}
\usepackage[all]{xy}
\usepackage{graphicx}
\usepackage{amsmath}
\usepackage{amssymb}
\usepackage{float}
\usepackage{array}
\usepackage{tikz}
\usepackage{mathtools}
\usepackage{mathrsfs} 
\usepackage{bbm}
\usepackage{hyperref}
\usepackage{nicefrac}
\usepackage{natbib}
\usepackage{comment}
\usepackage{mathbbol}
\usepackage{soul}
\DeclareMathOperator{\Tr}{Tr}
\def\bz{{\overline{z}}}
\usetikzlibrary{arrows.meta}

\arxivnumber{2607.17246} % if you have one

\title{\boldmath Non-Abelian Thirring model at large $N$}

\author[a,b]{Songyuan Li,}
\author[b]{Konstantinos Siampos}
\affiliation[a]{School of Physics, Sun Yat-sen University, Guangzhou 510275, China}
\affiliation[b]{Laboratory of Theoretical Physics, School of Physics,\\
Aristotle University of Thessaloniki, 54124 Thessaloniki, Greece}

\emailAdd{lisy353@mail.sysu.edu.cn}
\emailAdd{ksiampos@auth.gr}

\abstract{We consider the non-Abelian bosonized Thirring model for a semi-simple group $G$ at level $k$, with deformation parameter $\lambda$. We compute the two-point correlation functions of current and composite current operators to cubic order in $\lambda$, assuming large values of the quadratic Casimir $c_G$ of the group $G$ in the adjoint representation. From these, we extract the $\beta$-function and the anomalous dimensions of both the current and composite current operators, showing the absence of an additional critical point of order $\nicefrac{k}{c_G}$ to quartic order in $\lambda$. At this order, our findings agree with those of Destri and de Vega, but contradict the claim of Dashen and Frishman for the Fermionic non-Abelian Thirring model.
}

\begin{document}
\maketitle
\flushbottom

\section{Introduction and motivation}
\label{sec:intro}

Obtaining exact results in strongly interacting quantum field theories (QFTs) is a rather formidable task. In the last decades, important progress in understanding the strong coupling regime of QFTs has been achieved using the AdS/CFT duality, initiated by the example of the ${\cal N}=4$ SYM which is dual to the IIB supergravity on the $\text{AdS}_5\times S^5$ background~\cite{Maldacena:1997re}. A crucial ingredient in the study of the prototype, as well as its extensions, is integrability, considered from the sides of both gauge theory and string theory. Integrability has also played a pivotal r\^ole in the study of two-dimensional cases such as the non-Abelian Fermionic Thirring model~\cite{Dashen:1974gu,Karabali:1988sz,deVega:1983gn}, the Gross--Neveu model~\cite{Gross:1974jv} as well as the $T\bar T$ deformations~\cite{Zamolodchikov:2004ce,Smirnov:2016lqw}.

In the present work we are interested in studying the large $N$ limit of the bosonized non-Abelian Thirring model, given by the WZW action deformed by a classically marginal operator bilinear in the currents (see for example \cite{Kutasov:1989dt}). Its action is given by
\begin{equation}
\label{Thirring.Lor}
S=S_{\text{WZW},k}(g)-\frac{\lambda}{\pi}\int d^2\sigma J_+^a J_-^a\,,
\end{equation}
where in the above, the WZW action at level $k$ reads
\begin{equation}
S_{\text{WZW},k}(g)=-\frac{k}{2\pi}\int_{\partial B} d^2\sigma\,\text{Tr}\left(g^{-1}\partial_+ g g^{-1}\partial_-g\right)
+\frac{k}{{12}\pi}\int_B\text{Tr}(g^{-1}\text{d}g)^3
\end{equation}
and
\begin{equation}
J_+^a=i\sqrt{k}\,\text{Tr}(t_a\partial_+g g^{-1})\,,\, J_-^a=-i\sqrt{k}\,\text{Tr}(t_ag^{-1}\partial_-g)\,,\quad
\sigma^\pm=\tau\pm\sigma\,,\quad d^2\sigma=d\tau\, d\sigma\,,
\end{equation}
for a generic semi-simple Lie group $G$ with $g\in G$ which are parameterized by $X^\mu$, $\mu=1,2,\dots,\text{dimG}$. The corresponding Lie algebra is spanned by the generators $t_a$, which obey the commutation relations $[t_a,t_b]=f_{abc}t_c$. The matrices $t_a$ are traceless and are normalized as $\text{Tr}(t_at_b)=\delta_{ab}$. They are taken to be Hermitian and the structure constants $f_{abc}$ are imaginary and square to the quadratic Casimir $c_G$ in the adjoint representation, i.e. $f_{acd}f_{bcd}=-c_G\delta_{ab}$. For example if $G=SU(N)$, $c_G=2N$.

We are interested in exploring \eqref{Thirring.Lor} in a perturbative expansion around the WZW point and at first we will pass in the Euclidean regime $\tau\to-i t$, where
\begin{equation}
\sigma^+\to-i z\,,\quad \sigma^-\to-i\bar z\,,\quad
z=t+i\sigma\,,\quad \bar z=t-i\sigma\,,\quad
\partial_+\to i\partial\,,\quad \partial_-\to i\bar\partial\,.
\end{equation}
Working in the Euclidean, the action appears as $e^{-S_E}$ in the path integral, where
\begin{equation}
\label{Thirring.Eucl}
\begin{split}
&S_E = S^E_{\text{WZW},k}(g)+\frac{\lambda}{\pi}\int d^2z\, {O}(z,\bar z)\,,\quad {O}(z,\bar z):=J^a(z)\bar J^a(\bar z)\,,\\
&S^E_{\text{WZW},k}(g)=-\frac{k}{2\pi}\int_{\partial B} d^2z\,\text{Tr}\left(g^{-1}\partial g g^{-1}\bar\partial g\right)
-\frac{ik}{12\pi}\int_B\text{Tr}(g^{-1}\text{d}g)^3\,.
\end{split}
\end{equation}
In the above, we defined $d^2z=dt d\sigma$ and the currents 
$J=-\sqrt{k}\partial g g^{-1}$, $\bar J=\sqrt{k}g^{-1}\bar\partial g$ are chiral/antichiral $\bar\partial J=0=\partial\bar J$, obeying
the same algebra at level $k$~\cite{Witten,Knizhnik.Zamolodchikov}
\begin{equation}
\label{current.OPEs}
\begin{split}
&J^a(z_1)J^b(z_2)=\frac{\delta_{ab}}{z_{12}^2}+\frac{f_{abc} J^c(z_2)}{\sqrt{k}\,z_{12}}\,,\quad 
\bar J^a(\bar z_1)\bar J^b(\bar z_2)=\frac{\delta_{ab}}{\bar z_{12}^2}+\frac{f_{abc}\bar J^c(\bar z_2)}{\sqrt{k}\,\bar z_{12}}\,,\\
&J^a(z_1)\bar J^b(\bar z_2)=0\,,\quad z_{12}:=z_1-z_2
\end{split}
\end{equation}
and we recall that $f_{abc}$ scales as $\sqrt{c_G}$.

Assuming that $0\leqslant\lambda<1$, the quantum behavior of the above action \eqref{Thirring.Eucl} in the semi-classical regime has been studied to all-orders in $\lambda$ but to leading order in $\nicefrac{c_G}{k}\ll1$~\cite{Kutasov:1989dt}, see also~\cite{Gerganov:2000mt}\footnote{The all-loop beta function in Eq.~(18) of \cite{Gerganov:2000mt} is valid to leading order, $\nicefrac{c_G}{k} \ll 1$. At subleading order in $\nicefrac{c_G}{k}$, it already receives corrections -- see, for example, Eq.~(3.1) in \cite{Georgiou:2019nbz} or Eq.~(2.26) in \cite{Hoare:2019mcc}.}
\begin{equation}
\label{beta.Kutasov}
\beta^\lambda:=\frac{d\lambda}{d\ln\mu^2}=-\frac{c_G}{2k}\frac{\lambda^2}{(1+\lambda)^2}\leqslant0\,,
\end{equation}
where $\mu$ is the energy scale. Hence, we find that the non-Abelian Thirring model is flowing from the UV WZW fixed point at $\lambda=0$ towards the IR as $\lambda\to1$, describing a strongly coupled model.
Moreover, the anomalous dimension of the driving operator ${ O}$ is given by~\cite{Georgiou:2015nka}
\begin{equation}
\label{anomalous.O}
\gamma_{O}=-\frac{2c_G\lambda}{k}\frac{1-\lambda+\lambda^2}{(1-\lambda)(1+\lambda)^3}\leqslant0\,,
\end{equation}
hence $O$ is a marginally relevant operator.

Using path integral techniques it was argued in \cite{Kutasov:1989aw} that \eqref{Thirring.Eucl} possesses a non-perturbative symmetry in the parameter space of $(\lambda,k)$
\begin{equation}
\label{symmetry.general}
\lambda\to\lambda^{-1}\,,\quad k\to-k-c_G\,,
\end{equation}
which for $\nicefrac{c_G}{k}\ll1$, simplifies to $\lambda\to\lambda^{-1}$ and $k\to-k$\,.

In addition, the all-loop effective action of \eqref{Thirring.Lor} to order $\nicefrac{c_G}{k}\ll1$ is given by~\cite{Georgiou:2016iom}
\begin{equation}
\label{effective.action}
S_{\lambda,k}(g)=S_{\text{WZW},k}(g)-\frac{1}{\pi}\int d^2\sigma \left(\lambda^{-1}\mathbb{1}-D^T\right)_{ab}^{-1} J_+^a J_-^b\,,\quad 0\leqslant\lambda<1\,,
\end{equation}
where $D$ is the adjoint action and its matrix elements are given by $D_{ab}=\text{Tr}(t_agt_bg^{-1})$. This action is known as the $\lambda$-deformed model~\cite{Sfetsos:2013wia} and its one-loop RG flow $\alpha'\sim\nicefrac{c_G}{k}\ll1$ was worked out in~\cite{Itsios:2014lca}, which coincides with \eqref{beta.Kutasov}.

The effective action \eqref{effective.action} enjoys the explicit weak-strong duality 
\begin{equation}
\label{symmetry.action}
S_{\lambda,k}(g)=S_{\lambda^{-1},-k}(g^{-1})\,,
\end{equation}
thereby realizing the non-perturbative symmetry proposed in~\cite{Kutasov:1989aw}, for the non-Abelian bosonized Thirring model. In addition, the effective action enjoys two non-trivial zoom-in limits as $\lambda\to\pm1$ and $k\to\infty$~[\citenum{Sfetsos:2013wia}, \citenum{Georgiou:2016iom}],
\begin{equation}
\label{zoomin}
\begin{split}
&g=\mathbb{1}+i\frac{v_a t_a}{k}\,,\quad \lambda=1-\frac{\kappa^2}{k}\,,\quad k\to\infty\,,\\
&g=-\mathbb{1}+i\frac{v_a t_a}{k^{\nicefrac13}}\,,\quad \lambda=-1+\frac{1}{(b^2k)^{\nicefrac13}}\,,\quad k\to\infty\,,
\end{split}
\end{equation}
corresponding to the non-Abelian T-dual of the PCM and the pseudo-chiral model respectively. For $\lambda > 0$ and $k\gg c_G$, the non-Abelian bosonized Thirring model flows from the WZW model at the UV fixed point $\lambda=0$ toward the IR non-Abelian T-dual of the PCM as $\lambda\to1$.

Furthermore, using conformal perturbation, the symmetry \eqref{symmetry.action} and the zoom-in limits \eqref{zoomin}, 
we can compute anomalous dimensions of currents, of the composite operator $O$ and of affine primaries to 
leading order in  $c_G\ll k$ and all-orders in $\lambda$~\cite{Georgiou:2015nka,Georgiou:2016iom}. For example, the current anomalous dimension is found to be
\begin{equation}
\gamma_J=\frac{c_G}{k}\frac{\lambda^2}{(1-\lambda)(1+\lambda)^3}\,.
\end{equation}
whereas the $\beta$-function and the anomalous dimension of the composite operator are given by \eqref{beta.Kutasov} and \eqref{anomalous.O} respectively. 
In addition, using the leading-order results in $c_G\ll k$ and to all orders in $\lambda$ for the two- and three-point current correlation functions, 
the equal-time commutators of the currents were obtained and found to align with the underlying structure of the integrable $\lambda$-deformed $\sigma$-models~[\citenum{Sfetsos:2013wia}, \citenum{Georgiou:2016iom}]. Finally, extensions of the above include computing anomalous dimensions of operators built from an arbitrary number of currents of same chirality as well as mixed chirality, i.e. $JJ\bar J$~\cite{Georgiou:2019jcf}.

The objective of the present work is to extend the above construction to the regime $c_G\gg k$ and search for non-trivial fixed points, following the approach of Dashen and Frishman for the non-Abelian Fermionic Thirring model~\cite{Dashen:1974gu}. In this limit, the effective action \eqref{effective.action} is inapplicable; consequently, the zoom-in limits at $\lambda\to\pm1$ and the symmetry \eqref{symmetry.general} no longer hold. Therefore, the only viable tool is conformal perturbation theory. Within this framework, we will show that the effective 't Hooft coupling is given by $\Tilde{\lambda} = \frac{c_G}{k}\lambda$ and that no additional critical point of order $\nicefrac{k}{c_G}$ exists up to quartic order in $\tilde\lambda$, a result that aligns with the findings of Destri and de Vega~\cite{Destri,Destri.Vega} for the Fermionic variant.\footnote{See also the works of~\cite{Bondi:1989nq} and~\cite{Bennett:1999he}, for the renormalization of the non-Abelian Fermionic Thirring model at two- and three-loop respectively.}

This work is organized as follows: In section~\ref{Section:2}, we set-up the correlators at the CFT point $\lambda=0$ and away from it. In sections~\ref{Section.Composite} and~\ref{Section.currents}, we compute correlation functions of the driving operator and of the current operators respectively to cubic order in $\lambda$. In section~\ref{Section:Anomalous.Beta}, we evaluate the $\beta$-function to quartic order in $\lambda$ and leading order in $\nicefrac{c_G}{k}$. In section~\ref{Section:Concl.Out}, we conclude and discuss future directions. Finally, appendices~\ref{appendix:A}, \ref{appendixB}, \ref{appendixC}, \ref{appendixD} and \ref{appendixE} contain supporting material to the main text.

\section{Background}
\label{Section:2}

In this section, we lay out the computation, starting from the correlators at the CFT point and then prescribing the conformal perturbation theory and the necessary regularization scheme.

\subsection{Correlation functions at the CFT point}

We are interested in computing correlation functions of currents and our starting point will be the OPEs in \eqref{current.OPEs}. In our correlation functions we will need the one-point which vanishes as well as the two and three-point correlation functions
\begin{equation}
\langle J^{a_1}(z_1)J^{a_2}(z_2)\rangle = \frac{\delta_{a_1a_2}}{z_{12}^2}\label{2pt}
\end{equation}
and
\begin{equation}
 \langle J^{a_1}(z_1)J^{a_2}(z_2)J^{a_3}(z_3)\rangle = \frac{f_{a_1a_2a_3}}{\sqrt{k}\, z_{12}z_{13}z_{23}}\,. \label{3pt}
\end{equation}
Using the Ward identity
\begin{eqnarray}
&&\langle J^a(z) J^{a_1}(z_1) J^{a_2}(z_2) \cdots J^{a_n}(z_n) \rangle
 = \frac{1}{\sqrt{k}} \sum_{i=1}^n \frac{f_{a a_i b_i}}{z - z_i} \langle J^{a_1}(z_1) J^{a_2}(z_2) \cdots J^{b_i}(z_i) \cdots J^{a_n}(z_n) \rangle \nonumber \\
& & \qquad\qquad\qquad + \sum_{i=1}^n \frac{\delta_{a a_i}}{(z - z_i)^2} \langle J^{a_1}(z_1) J^{a_2}(z_2) \cdots J^{a_{i-1}}(z_{i-1}) J^{a_{i+1}}(z_{i+1}) \cdots J^{a_n}(z_n) \rangle \,,\label{Wardidentity}
\end{eqnarray}
we can work out the current four-point which to leading order in $\nicefrac{c_G}{k}$ reads
\begin{equation}
\langle J^{a_1}(z_1)J^{a_2}(z_2)J^{a_3}(z_3)J^{a_4}(z_4)\rangle = \frac{1}{k} 
 \left(\frac{f_{a_1a_3e}f_{a_2a_4e}}{z_{12}z_{13}z_{24}z_{34}}-\frac{f_{a_1a_4e}f_{a_2a_3e}}{z_{12}z_{14}z_{23}z_{34}}\right)\,,\label{4pt}
\end{equation}
as well as the five-point correlation function to leading order in $\nicefrac{c_G}{k}$
\begin{equation}
\begin{split}
   &\langle J^{a_1}(z_1)J^{a_2}(z_2)J^{a_3}(z_3)J^{a_4}(z_4)J^{a_5}(z_5)\rangle\\
   &=\frac{1}{k^{\frac{3}{2}}}\left(-\frac{f_{a_1a_4c}f_{a_2ce}f_{a_3a_5e}}{z_{12}z_{14}z_{23}z_{35}z_{45}}-\frac{f_{a_1a_5c}f_{a_2a_4e}f_{a_3ce}}{z_{13}z_{15}z_{23}z_{24}z_{45}}+\frac{f_{a_1ce}f_{ca_2a_4}f_{a_3a_5e}}{z_{12}z_{13}z_{24}z_{35}z_{45}}\right.\\
   &\quad\quad\left. +\frac{f_{a_1a_5c}f_{a_2ce}f_{a_3a_4e}}{z_{12}z_{15}z_{23}z_{34}z_{45}}+\frac{f_{a_1a_4c}f_{a_2a_5e}f_{a_3ce}}{z_{13}z_{14}z_{23}z_{25}z_{45}}-\frac{f_{a_1ce}f_{ca_2a_5}f_{a_3a_4e}}{z_{12}z_{13}z_{25}z_{34}z_{45}}\right)\,.\label{eq:5ptfunction}
   \end{split}
\end{equation}

\subsection{Conformal perturbation theory}

Working in Euclidean the action appears in the path integral as $e^{-S_E}$, where $S_E$ is the non-Abelian bosonized 
Thirring model~\eqref{Thirring.Eucl}. A generic correlation function is given by
\begin{equation}
\langle F_1(x_1,\bar x_1)F_1(x_2,\bar x_2)\cdots\rangle_{\lambda}=
\left\langle F_1(x_1,\bar x_1)F_1(x_2,\bar x_2)\cdots\exp\left\{-\frac{\lambda}{\pi}\int d^2z\, {O}(z,\bar z)\right\}\right\rangle\,,
\end{equation}
hence the $n^\text{th}$-order contribution reads
\begin{equation}
\label{n.correlator}
\begin{split}
\langle F_1(x_1,\bar x_1)F_1(x_2,\bar x_2)\cdots\rangle^{(n)}_\lambda=
\frac{1}{n!}\left(-\frac\lambda\pi\right)^n&\int d^2z_{1,2,\cdots, n}\langle 
F_1(x_1,\bar x_1)F_1(x_2,\bar x_2)\cdots\\
&J^{a_1}(z_1)J^{a_2}(z_2)\cdots 
\bar J^{a_1}(\bar z_1)\bar J^{a_2}(\bar z_2)\cdots
\rangle\,,
\end{split}
\end{equation}
where $d^2z_{1,2,\cdots, n}=d^2z_1 d^2z_2\cdots d^2z_n$. Therefore we will face multiple integrals, some of which might require regularization. 
Our prescription is as follows:
\begin{itemize}
\item
The domain of integration $\mathcal{D}$ is a disc of radius $R$, containing all the external (non-integrated) points $x_i$, as read in the correlator \eqref{n.correlator},  such that $R\gg|x_i|$, see figure below, where $\Re$ and $\Im$ denotes the real and imaginary part respectively.
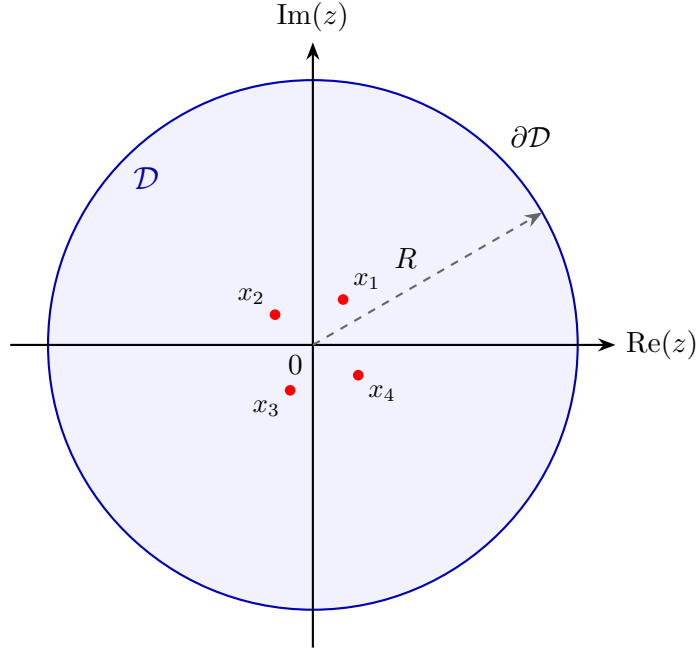
\begin{figure}[H]
    \centering
\begin{tikzpicture}
    % Define the radius of the disc
    \def\R{3.5}
    
    % 1. Draw and shade the domain of integration (Disc D)
    \fill[blue!5] (0,0) circle (\R);
    \draw[thick, blue!70!black] (0,0) circle (\R);
    
    % 2. Draw the Axes (Real and Imaginary)
    \draw[->, >=Stealth, thick] (-\R - 0.5, 0) -- (\R + 0.5, 0) node[right] {$\Re(z)$};
    \draw[->, >=Stealth, thick] (0, -\R - 0.5) -- (0, \R + 0.5) node[above] {$\Im(z)$};
    
    % 3. Label the domain and its boundary
    \node[blue!70!black, font=\large] at (-2.2, 2.2) {$\mathcal{D}$};
    \node[anchor=south west] at (45:\R) {$\partial\mathcal{D}$};
    
    % 4. Draw the radius R (at a 30-degree angle to stay clear of the axes)
    \draw[dashed, ->, >=Stealth, thick, gray!80!black] (0,0) -- (30:\R) node[midway, above left, black] {$R$};
    
    % Origin / Center
    \node[below left] at (0,0) {$0$};
    
    % 5. Cluster of external points z_i close to the center (R >> |z_i|)
    \fill[red] (0.4, 0.6) circle (2pt) node[above right, black, font=\small] {$x_1$};
    \fill[red] (-0.5, 0.4) circle (2pt) node[above left, black, font=\small] {$x_2$};
    \fill[red] (-0.3, -0.6) circle (2pt) node[below left, black, font=\small] {$x_3$};
    \fill[red] (0.6, -0.4) circle (2pt) node[below right, black, font=\small] {$x_4$};
    
    % Optional descriptive caption text below the plot
    \node[below, align=center, yshift=-0.8cm] at (0, -\R) {
       % Domain of integration $\mathcal{D}$ with radius $R$ containing all points $z_i$ such that $R \gg |z_i|$.
    };
\end{tikzpicture}
\caption{Domain of integration, a disc of radius $R$, containing all the external points $x_i$.}
\end{figure}
\item
In a multiple integral $d^2z_{1,2,\cdots, n}$, we choose to integrate starting \underline{from right to left} and we \underline{never alter} the integration order.
\item
We will keep all possible Dirac delta functions as well as derivatives of them that appear after integration, see below Eq.~\eqref{basic.delta}.
\item
In divergent integrals, whenever two points of any nature, either external or internal, coincide we will use  a small distance regulator $\epsilon$.

\end{itemize}

\noindent
To proceed we will introduce the basic integral
\begin{equation}
\label{int.basic}
\int\frac{d^2z}{(z-z_1)(\bar z_2-\bar z)}=\pi\ln\frac{|z_{12}|^2}{R^2}\,,
\end{equation}
which can be obtained at least in two-ways, either by direct integration or through Stokes theorem in complex coordinates, a proof which we revisit in appendix~\ref{appendix:A}.
If the external points in \eqref{int.basic} coincide, we use $\epsilon$ as a UV regulator 
\begin{equation}
\int\frac{d^2z}{(z-z_1)(\bar z_1-\bar z)}=\pi\ln\frac{\epsilon^2}{R^2}\,.
\end{equation}
Starting from \eqref{int.basic} we may also compute
\begin{equation}
\int\frac{d^2z}{(z-z_1)^2(\bar z-\bar z_2)}=-\frac{\pi}{z_{12}}\,,\quad
\int\frac{d^2z}{(z-z_1)(\bar z-\bar z_2)^2}=\frac{\pi}{\bar z_{12}}\,,
\end{equation}
as well as
\begin{equation}
\label{basic.delta}
\int\frac{d^2z}{(z-z_1)^2(\bar z-\bar z_2)^2}=\pi^2\delta^{(2)}(z_{12})\,.
\end{equation}
In appendix~\ref{appendix:A} we work out several integrals that are required in evaluating the correlation functions of the current and composite operators.

\section{Composite operator}
\label{Section.Composite}

The objective of this section is to compute the leading contribution in $\nicefrac{c_G}{k}$ up to order $\lambda^3$ for the correlation functions of the driving composite operator $O=J^a(z)\bar J^a(\bar z)$.

\subsection{A 't Hooft limit}
\label{Section.Composite.tHooft}

At $L$ loop the correction to the $OO$ correlator is given by
\begin{eqnarray}
    &&\frac{(-\lambda)^L}{L!\pi^L}\int d^2 z_{3,4,\cdots,L+2}\langle J^{a_1}(z_1)J^{a_2}(z_2)J^{a_3}(z_3)\cdots J^{a_{L+2}}(z_{L+2})\rangle\nonumber\\
    &&\qquad\qquad\times\langle \Bar{J}^{a_1}(\bz_1)\Bar{J}^{a_2}(\bz_2)\Bar{J}^{a_3}(\bz_3)\cdots\Bar{J}^{a_{L+2}}(\bz_{L+2})\rangle\,.
\end{eqnarray}
When $c_G\gg k$, we keep only the leading orders in $\nicefrac{c_G}{k}$. This means that we keep only the second term in \eqref{current.OPEs} when doing an OPE unless its contribution vanishes. We also know that the structure constants are of order $\sqrt{c_G}$, so by Ward identity \eqref{Wardidentity} we know that the $L+3$-point function of $J$ is of order $\sqrt{\nicefrac{c_G}{k}}$ times the $L+2$-point function of $J$ for $L\geqslant 0$. Knowing that for $L=0$ we have the two-point function being of order $\left(\nicefrac{c_G}{k}\right)^0$, we find that the $L+2$-point function of $J$ is of order $\left(\nicefrac{c_G}{k}\right)^{\frac{L}{2}}$. Therefore the $L$ loop correction to the $OO$ correlator $\langle O(z_1)O(z_2)\rangle$ is of the form
\begin{equation}
    \left(\lambda\frac{c_G}{k}\right)^L\times I_L(z_1,z_2)\,,
\end{equation}
where $I_L(z_1,z_2)$ is some integral whose result only depends on $z_{1,2}$ (and the UV cutoff $\epsilon$). This motivates us to consider a 't Hooft limit where 
$\lambda\to0$, $\nicefrac{c_G}{k}\to\infty$ while
\begin{equation}
\label{lambda.tHooft}
    \Tilde{\lambda} = \frac{c_G}{k}\lambda\,,
\end{equation}
is kept fixed and we consider a perturbation expansion with respect to $\Tilde{\lambda}$.
\subsection{Tree level and 1-loop}
Using the OPE algebra \eqref{current.OPEs} we can easily compute the correlator at tree-level:
\begin{equation}
    \langle O(z_1)O(z_2)\rangle^{(0)} = \frac{\dim G}{|z_{12}|^4}\,.
\end{equation}
Using the OPE algebra \eqref{current.OPEs} and \eqref{int.basic} we can compute the 1-loop correction~\cite{Georgiou:2015nka}\footnote{Note that the sign in Eq. (3.7) of \cite{Georgiou:2015nka} is incorrect; the correct expression can be found by taking $k_L=k_R=k$ in Eq. (2.18) of \cite{Georgiou:2016zyo}.}:
\begin{equation}
    \langle O(z_1)O(z_2)\rangle^{(1)} = -\frac{2\dim G\,\lambda\, c_G}{k|z_{12}|^4} \ln\frac{\epsilon^2}{|z_{12}|^2}\,.
\end{equation}
\subsection{2-loop}
\label{Section.2loop.composite}
We now turn to the 2-loop correction to the correlator $\langle O(z_1)O(z_2)\rangle$ which is given by
\begin{equation}
    \langle O(z_1)O(z_2)\rangle^{(2)} = \frac{(-\lambda)^2}{2!\pi^2}\int d^2 z_{3,4} \langle J^{a_1}(z_1)J^{a_2}(z_2)J^{a_3}(z_3)J^{a_4}(z_4)\rangle \langle \Bar{J}^{a_1}(\bz_1)\Bar{J}^{a_2}(\bz_2)\Bar{J}^{a_3}(\bz_3)\Bar{J}^{a_4}(\bz_4)\rangle\,.
\end{equation}
Using the four-point functions \eqref{4pt}, this is equal to
\begin{equation}
\begin{split}
    &\frac{(-\lambda)^2}{2!\pi^2 k^2}\int d^2 z_{3,4} \left(\frac{f_{a_1a_3c}f_{a_2a_4c}}{z_{12}z_{13}z_{24}z_{34}}-\frac{f_{a_1a_4c}f_{a_2a_3c}}{z_{12}z_{14}z_{23}z_{34}}\right)
    \left(\frac{f_{a_1a_3e}f_{a_2a_4e}}{\bz_{12}\bz_{13}\bz_{24}\bz_{34}}-\frac{f_{a_1a_4e}f_{a_2a_3e}}{\bz_{12}\bz_{14}\bz_{23}\bz_{34}}\right)\\
    &= \frac{(-\lambda)^2}{2!\pi^2 k^2}\int d^2 z_{3,4} \left(\frac{f_{a_1a_3c}f_{a_2a_4c}f_{a_1a_3e}f_{a_2a_4e}}{z_{12}z_{13}z_{24}z_{34}\bz_{12}\bz_{13}\bz_{24}\bz_{34}}-\frac{f_{a_1a_3c}f_{a_2a_4c}f_{a_1a_4e}f_{a_2a_3e}}{z_{12}z_{13}z_{24}z_{34}\bz_{12}\bz_{14}\bz_{23}\bz_{34}}\right.\\
    &\quad\quad\quad\quad\quad\quad \left.-\frac{f_{a_1a_4c}f_{a_2a_3c}f_{a_1a_3e}f_{a_2a_4e}}{z_{12}z_{14}z_{23}z_{34}\bz_{12}\bz_{13}\bz_{24}\bz_{34}}+\frac{f_{a_1a_4c}f_{a_2a_3c}f_{a_1a_4e}f_{a_2a_3e}}{z_{12}z_{14}z_{23}z_{34}\bz_{12}\bz_{14}\bz_{23}\bz_{34}}\right)\\
    &=\frac{(-\lambda)^2c_G^2\dim G}{2!\pi^2 k^2}\int d^2 z_{3,4} \left(\frac{1}{z_{12}z_{13}z_{24}z_{34}\bz_{12}\bz_{13}\bz_{24}\bz_{34}}-\frac{1}{2z_{12}z_{13}z_{24}z_{34}\bz_{12}\bz_{14}\bz_{23}\bz_{34}}\right.\\
    &\quad\quad\quad\quad\quad\quad\quad\quad \left.-\frac{1}{2z_{12}z_{14}z_{23}z_{34}\bz_{12}\bz_{13}\bz_{24}\bz_{34}}+\frac{1}{z_{12}z_{14}z_{23}z_{34}\bz_{12}\bz_{14}\bz_{23}\bz_{34}}\right)\,,
    \end{split}
\end{equation}
where we used results of appendix~\ref{appendixB} to contract the structure constants.
The first line in the result is related to the second line by interchanging $z_1$ and $z_2$. As we will show below that after integrating each term will give a result invariant under interchanging $z_1$ and $z_2$, we can further write
\begin{equation}
\label{composite.twoloop}
    \langle O(z_1)O(z_2)\rangle^{(2)} = \frac{(-\lambda)^2c_G^2\dim G}{2!\pi^2 k^2 |z_{12}|^2}\int d^2 z_{3,4} 
    \left(\frac{2}{z_{13}z_{24}z_{34}\bz_{13}\bz_{24}\bz_{34}}-\frac{1}{z_{13}z_{24}z_{34}\bz_{14}\bz_{23}\bz_{34}}\right)\,.
\end{equation}
The first integral that one needs to compute is
\begin{equation}
\label{composite.twoloop1}
   2 \int  \frac{d^2 z_{3,4}}{z_{13}z_{24}z_{34}\bz_{13}\bz_{24}\bz_{34}}\,.
\end{equation}
Performing the $z_4$ integration first we find it equal to
\begin{equation}
\begin{split}
    & -4\pi \int  \frac{d^2 z_3}{z_{13}z_{23}\bz_{13}\bz_{23}} \ln\frac{\epsilon^2}{|z_{23}|^2}\\
    &= \frac{4\pi}{|z_{12}|^2} \int d^2 z_3 \left(\frac{1}{z_{13}}-\frac{1}{z_{23}}\right)\left(\frac{1}{\bz_{13}}-\frac{1}{\bz_{23}}\right)\left(-\ln\frac{\epsilon^2}{R^2}+\ln\frac{|z_{23}|^2}{R^2}\right)\\
    &= -\frac{4\pi}{|z_{12}|^2}\ln\frac{\epsilon^2}{R^2}\int d^2 z_3 \left(\frac{1}{z_{13}\bz_{13}}-\frac{1}{z_{13}\bz_{23}}-\frac{1}{z_{23}\bz_{13}}+\frac{1}{z_{23}\bz_{23}}\right)\\
    &+\frac{4\pi}{|z_{12}|^2}\int d^2 z_3 \left(\frac{1}{z_{13}\bz_{13}}-\frac{1}{z_{13}\bz_{23}}-\frac{1}{z_{23}\bz_{13}}+\frac{1}{z_{23}\bz_{23}}\right)\ln\frac{|z_{23}|^2}{R^2}\,.
\end{split}
\end{equation}
One can then do the $z_3$ integral using \eqref{spin-off.I} and \eqref{Iwith2equal3} and find it equal to
\begin{equation}
    \frac{8\pi^2}{|z_{12}|^2}\ln\frac{\epsilon^2}{R^2}\ln\frac{\epsilon^2}{|z_{12}|^2}-\frac{4\pi^2}{|z_{12}|^2}\ln\frac{\epsilon^2}{R^2}\ln\frac{|z_{12}|^2}{R^2}+\frac{6\pi^2}{|z_{12}|^2}\ln^2\frac{|z_{12}|^2}{R^2}-\frac{2\pi^2}{|z_{12}|^2}\ln^2\frac{\epsilon^2}{R^2}\,,
\end{equation}
which simplifies to
\begin{equation}
\label{composite.twoloop2}
     \frac{6\pi^2}{|z_{12}|^2}\ln^2\frac{\epsilon^2}{|z_{12}|^2}\,.
\end{equation}
The second integral that one needs to compute is
\begin{equation}
    \int  \frac{d^2 z_{3,4}}{z_{13}z_{24}z_{34}\bz_{14}\bz_{23}\bz_{34}}\,,
\end{equation}
we integrate over $z_4$ first and find it equal to
\begin{equation}
    \pi\int \frac{d^2 z_3 }{z_{13}z_{23}\bz_{13}\bz_{23}} 
    \left(-\ln\frac{|z_{12}|^2}{R^2}+\ln\frac{|z_{13}|^2}{R^2}+\ln\frac{|z_{23}|^2}{R^2}-\ln\frac{\epsilon^2}{R^2}\right)\,.
\end{equation}
One can then do the $z_3$ integral, again using \eqref{spin-off.I} and \eqref{Iwith2equal3}, and find it equal to 
\begin{equation}
    \frac{2\pi^2}{|z_{12}|^2}\ln\frac{\epsilon^2}{|z_{12}|^2}\left(\ln\frac{|z_{12}|^2}{R^2}+\ln\frac{\epsilon^2}{R^2}\right)-\frac{2\pi^2}{|z_{12}|^2}\ln\frac{\epsilon^2}{R^2}\ln\frac{|z_{12}|^2}{R^2}+\frac{3\pi^2}{|z_{12}|^2}\ln^2\frac{|z_{12}|^2}{R^2}-\frac{\pi^2}{|z_{12}|^2}\ln^2\frac{\epsilon^2}{R^2}\,,
\end{equation}
which simplifies to
\begin{equation}
    \frac{\pi^2}{|z_{12}|^2}\ln^2\frac{\epsilon^2}{|z_{12}|^2}\,.
\end{equation}
So in the end we find from \eqref{composite.twoloop}
\begin{equation}
    \langle O(z_1)O(z_2)\rangle^{(2)} = \frac{5\lambda^2c_G^2\dim G}{2 k^2 |z_{12}|^4}\ln^2\frac{\epsilon^2}{|z_{12}|^2}\,.
\end{equation}
We note that there is an additional contribution to the two-loop term of order $\nicefrac{c_G}{k}$, found from Eq.(2.21) in \cite{Georgiou:2016zyo} for $k_L=k_R=k$, that is 
\begin{equation}
    \langle O(z_1)O(z_2)\rangle^{(2)}_\text{sub} = \frac{6\lambda^2c_G\dim G}{ k |z_{12}|^4}\ln\frac{\epsilon^2}{|z_{12}|^2}\,.
\end{equation}
However, this term is sub-leading when $c_G\gg k$.

\subsection{3-loop}
\label{OO3loopint}
Now let's take a look at the OO correlator at 3-loop, this is given by
\begin{equation}
\begin{split}
    \langle O(z_1)O(z_2)\rangle^{(3)} &= \frac{(-\lambda)^3}{3!\pi^3}\int d^2 z_{3,4,5} \langle J^{a_1}(z_1)J^{a_2}(z_2)J^{a_3}(z_3)J^{a_4}(z_4)J^{a_5}(z_5)\rangle\\
    & \qquad\qquad\qquad\,\,\times\langle \Bar{J}^{a_1}(\bz_1)\Bar{J}^{a_2}(\bz_2)\Bar{J}^{a_3}(\bz_3)\Bar{J}^{a_4}(\bz_4)\Bar{J}^{a_5}(\bz_5)\rangle\,.\label{OO3loop}
    \end{split}
\end{equation}
Using \eqref{eq:5ptfunction}, the leading order contribution when $\nicefrac{c_G}{k}\gg1$ is
\begin{equation}
\begin{split}
    &\langle O(z_1)O(z_2)\rangle^{(3)}\\
    &= \frac{(-\lambda)^3}{3!\pi^3k^3}\int d^2 z_{3,4,5} \left(-\frac{f_{a_1a_4c}f_{a_2ce}f_{a_3a_5e}}{z_{12}z_{14}z_{23}z_{35}z_{45}}-\frac{f_{a_1a_5c}f_{a_2a_4e}f_{a_3ce}}{z_{13}z_{15}z_{23}z_{24}z_{45}}+\frac{f_{a_1ce}f_{ca_2a_4}f_{a_3a_5e}}{z_{12}z_{13}z_{24}z_{35}z_{45}}\right.\\
    &\qquad\qquad\qquad\qquad\left.+\frac{f_{a_1a_5c}f_{a_2ce}f_{a_3a_4e}}{z_{12}z_{15}z_{23}z_{34}z_{45}}+\frac{f_{a_1a_4c}f_{a_2a_5e}f_{a_3ce}}{z_{13}z_{14}z_{23}z_{25}z_{45}}-\frac{f_{a_1ce}f_{ca_2a_5}f_{a_3a_4e}}{z_{12}z_{13}z_{25}z_{34}z_{45}}\right) \\
    & \qquad\qquad\qquad\quad\times\left(-\frac{f_{a_1a_4f}f_{a_2fg}f_{a_3a_5g}}{\bz_{12}\bz_{14}\bz_{23}\bz_{35}\bz_{45}}-\frac{f_{a_1a_5f}f_{a_2a_4g}f_{a_3fg}}{\bz_{13}\bz_{15}\bz_{23}\bz_{24}\bz_{45}}+\frac{f_{a_1fg}f_{fa_2a_4}f_{a_3a_5g}}{\bz_{12}\bz_{13}\bz_{24}\bz_{35}\bz_{45}}\right.\\
    &\qquad\qquad\qquad\qquad\left.+\frac{f_{a_1a_5f}f_{a_2fg}f_{a_3a_4g}}{\bz_{12}\bz_{15}\bz_{23}\bz_{34}\bz_{45}}+\frac{f_{a_1a_4f}f_{a_2a_5g}f_{a_3fg}}{\bz_{13}\bz_{14}\bz_{23}\bz_{25}\bz_{45}}-\frac{f_{a_1fg}f_{fa_2a_5}f_{a_3a_4g}}{\bz_{12}\bz_{13}\bz_{25}\bz_{34}\bz_{45}}\right)\,.
    \label{eq:OO3-loop}
    \end{split}
\end{equation}
This will give us in total 36 integrals which we will compute in detail in appendix \ref{appendixC}. There we find
\begin{equation}
   \langle O(z_1)O(z_2)\rangle^{(3)} = -\frac{5\lambda^3c_G^3\dim G}{2k^3|z_{12}|^4}\ln^3\frac{\epsilon^2}{|z_{12}|^2}
   +\frac{10\zeta(3)\lambda^3c_G^3\dim G}{3k^3|z_{12}|^4}\,,
\end{equation}
which is the leading contribution at order $\lambda^3$ for $\nicefrac{c_G}{k}\gg1$. At the same order there are also accompanying terms of order 
$\lambda^3\nicefrac{c_G}{k}$~\cite{Georgiou:2015nka} and $\lambda^3\nicefrac{c_G^2}{k^2}$, which are subleading in the 't Hooft limit.

\section{Current two-point correlators}
\label{Section.currents}

The objective of this section is to compute the leading contribution in $\nicefrac{c_G}{k}$ up to order $\lambda^3$ for the correlation functions of the current operators $J$ and $\bar J$.

\subsection{JJ}

The correction to the $JJ$ correlation function up to 2-loop has been computed in~\cite{Georgiou:2015nka}
\begin{equation}
\label{JJ.twoloop}
\langle J^{a_1}(z_1)J^{a_2}(z_2)\rangle=\frac{\delta_{a_1a_2}}{z_{12}^2}\left(1+\frac{c_G\,\lambda^2}{k}\ln\frac{\epsilon^2}{|z_{12}|^2}\right)\,.
\end{equation}
The 3-loop correction to the $JJ$ correlator is given by
\begin{equation}
    \langle J^{a_1}(z_1)J^{a_2}(z_2)\rangle^{(3)}=\frac{(-\lambda)^3}{3!\pi^3}\int d^2z_{3,4,5} \langle J^{a_1}(z_1)J^{a_2}(z_2)J^{a_3}(z_3)J^{a_4}(z_4)J^{a_5}(z_5)\rangle \langle\Bar{J}^{a_3}(\bz_3)\Bar{J}^{a_4}(\bz_4)\Bar{J}^{a_5}(\bz_5)\rangle\,.
\end{equation}
To the leading order in $\nicefrac{c_G}{k}$, we have
\begin{eqnarray}
    &&\langle J^{a_1}(z_1)J^{a_2}(z_2)J^{a_3}(z_3)J^{a_4}(z_4)J^{a_5}(z_5)\rangle \langle\Bar{J}^{a_3}(\bz_3)\Bar{J}^{a_4}(\bz_4)\Bar{J}^{a_5}(\bz_5)\rangle\\
    &=& \frac{1}{k^2\bz_{34}\bz_{35}\bz_{45}}\left(-\frac{f_{a_1a_4c}f_{a_2ce}f_{a_3a_5e}f_{a_3a_4a_5}}{z_{12}z_{14}z_{23}z_{35}z_{45}}-\frac{f_{a_1a_5c}f_{a_2a_4e}f_{a_3ce}f_{a_3a_4a_5}}{z_{13}z_{15}z_{23}z_{24}z_{45}}+\frac{f_{a_1ce}f_{ca_2a_4}f_{a_3a_5e}f_{a_3a_4a_5}}{z_{12}z_{13}z_{24}z_{35}z_{45}}\right.\nonumber\\ &&\quad\quad\quad\quad\quad\quad\left.+\frac{f_{a_1a_5c}f_{a_2ce}f_{a_3a_4e}f_{a_3a_4a_5}}{z_{12}z_{15}z_{23}z_{34}z_{45}}+\frac{f_{a_1a_4c}f_{a_2a_5e}f_{a_3ce}f_{a_3a_4a_5}}{z_{13}z_{14}z_{23}z_{25}z_{45}}-\frac{f_{a_1ce}f_{ca_2a_5}f_{a_3a_4e}f_{a_3a_4a_5}}{z_{12}z_{13}z_{25}z_{34}z_{45}}\right)\nonumber\,.
\end{eqnarray}
We then sum over the indices of the structure constant using the result in appendix \ref{appendixB} and find
\begin{equation}
\label{JJ3loopint}
\begin{split}
    &\langle J^{a_1}(z_1)J^{a_2}(z_2)\rangle^{(3)}\\
    &= \frac{(-\lambda)^3c_G^2\delta_{a_1a_2}}{3!\pi^3k^2} \int d^2 z_{3,4,5} \left( \frac{1}{z_{12}z_{13}z_{25}z_{34}z_{45}\bz_{34}\bz_{35}\bz_{45}}+\frac{1}{z_{12}z_{13}z_{24}z_{35}z_{45}\bz_{34}\bz_{35}\bz_{45}}\right.\\
    &\qquad\qquad\qquad\qquad\qquad\,\left. +\frac{1}{2z_{13}z_{14}z_{23}z_{25}z_{45}\bz_{34}\bz_{35}\bz_{45}}\right)+ (1\leftrightarrow 2)\,.
    \end{split}
\end{equation}
These three integrals are done in appendix \ref{appendixD} and we get the following result
\begin{equation}
    \langle J^{a_1}(z_1)J^{a_2}(z_2)\rangle^{(3)} = \frac{\lambda^3c_G^2\delta_{a_1a_2}}{k^2z_{12}^2}\left(-\frac{1}{2}\ln^2\frac{\epsilon^2}{|z_{12}|^2}+\ln\frac{\epsilon^2}{|z_{12}|^2}\right)\,.
\end{equation}
We note that there is an additional contribution to the three-loop term of order $\nicefrac{c_G}{k}$~\cite{Georgiou:2015nka} 
\begin{equation}
    \langle J^{a_1}(z_1)J^{a_2}(z_2)\rangle^{(3)}_\text{sub} = -\frac{2\lambda^3c_G\delta_{a_1a_2}}{k\,z_{12}^2}\ln\frac{e\,\epsilon^2}{|z_{12}|^2}\,.
\end{equation}
However, this term is sub-leading when $\nicefrac{c_G}{k}\gg1$.

\subsection{JJbar}
\label{Subsection:JJbar.twoloop}

The correction to the $J\bar J$ correlation function up to 2-loop has been computed in~\cite{Georgiou:2016iom}
\begin{equation}
\label{JJbar.twoloop}
\langle J^{a_1}(z_1)\Bar{J}^{a_2}(\bz_2)\rangle=-\pi\lambda\delta_{a_1a_2}\delta^{(2)}(z_{12})
-\frac{\lambda^2c_G\delta_{a_1a_2}}{k}\left(\frac{1}{|z_{12}|^2}+\pi\delta^{(2)}(z_{12})\left(1-\frac12\ln\frac{\epsilon^2}{|z_{12}|^2}\right)
\right)\,.
\end{equation}
We are interested in computing the 3-loop correction to the $J\Bar{J}$ correlator, this is given by
\begin{equation}
    \langle J^{a_1}(z_1)\Bar{J}^{a_2}(\bz_2)\rangle^{(3)}=\frac{(-\lambda)^3}{3!\pi^3}\int d^2z_{3,4,5} \langle J^{a_1}(z_1)J^{a_3}(z_3)J^{a_4}(z_4)J^{a_5}(z_5)\rangle \langle\Bar{J}^{a_2}(\bz_2)\Bar{J}^{a_3}(\bz_3)\Bar{J}^{a_4}(\bz_4)\Bar{J}^{a_5}(\bz_5)\rangle\,.\label{jjbar1}
\end{equation}
To the leading order in $\nicefrac{c_G}{k}$, we find it equal to
\begin{equation}
\begin{split}
    &\frac{(-\lambda)^3}{3!\pi^3 k^2}\int d^2z_{3,4,5} \left(\frac{f_{a_1a_4c}f_{a_3a_5c}}{z_{13}z_{14}z_{35}z_{45}}-\frac{f_{a_1a_5c}f_{a_3a_4c}}{z_{13}z_{15}z_{34}z_{45}}\right)\left(\frac{f_{a_2a_4e}f_{a_3a_5e}}{\bz_{23}\bz_{24}\bz_{35}\bz_{45}}-\frac{f_{a_2a_5e}f_{a_3a_4e}}{\bz_{23}\bz_{25}\bz_{34}\bz_{45}}\right)\\
    &= \frac{(-\lambda)^3}{3!\pi^3 k^2}\int d^2z_{3,4,5} \left(\frac{f_{a_1a_4c}f_{a_3a_5c}f_{a_2a_4e}f_{a_3a_5e}}{z_{13}z_{14}z_{35}z_{45}\bz_{23}\bz_{24}\bz_{35}\bz_{45}}-\frac{f_{a_1a_5c}f_{a_3a_4c}f_{a_2a_4e}f_{a_3a_5e}}{z_{13}z_{15}z_{34}z_{45}\bz_{23}\bz_{24}\bz_{35}\bz_{45}}\right.\\
    &\quad\quad\quad\quad\quad\quad\quad \left.-\frac{f_{a_1a_4c}f_{a_3a_5c}f_{a_2a_5e}f_{a_3a_4e}}{z_{13}z_{14}z_{35}z_{45}\bz_{23}\bz_{25}\bz_{34}\bz_{45}}+\frac{f_{a_1a_5c}f_{a_3a_4c}f_{a_2a_5e}f_{a_3a_4e}}{z_{13}z_{15}z_{34}z_{45}\bz_{23}\bz_{25}\bz_{34}\bz_{45}}\right)\,.
 \end{split}
\end{equation}
We then sum over the indices of the structure constant using the result in appendix \ref{appendixB} and find
\begin{equation}
\begin{split}
\label{JJbar3loopint}
    &\langle J^{a_1}(z_1)\Bar{J}^{a_2}(\bz_2)\rangle^{(3)}\\
    &=\frac{(-\lambda)^3c_G^2\delta_{a_1a_2}}{3!\pi^3 k^2}\int d^2z_{3,4,5} \left(\frac{1}{z_{13}z_{14}z_{35}z_{45}\bz_{23}\bz_{24}\bz_{35}\bz_{45}}-\frac{1}{2z_{13}z_{15}z_{34}z_{45}\bz_{23}\bz_{24}\bz_{35}\bz_{45}}\right.\\
    &\quad\quad\quad\quad\quad\quad\quad\quad\quad\left. -\frac{1}{2z_{13}z_{14}z_{35}z_{45}\bz_{23}\bz_{25}\bz_{34}\bz_{45}}+\frac{1}{z_{13}z_{15}z_{34}z_{45}\bz_{23}\bz_{25}\bz_{34}\bz_{45}}\right).
 \end{split}
\end{equation}
The four integrals are evaluated in appendix \ref{appendixE} and adding them up we get the following result
\begin{equation}
\begin{split}
    \langle J^{a_1}(z_1)\Bar{J}^{a_2}(\bz_2)\rangle^{(3)} &= -\frac{\lambda^3c_G^2\delta_{a_1a_2}}{k^2}\left[\frac{1}{|z_{12}|^2}\left(2-\ln\frac{\epsilon^2}{|z_{12}|^2}\right)\right.\\
    & \qquad\left.+ \frac{\pi}{4}\delta^{(2)}(z_{12})\left(6-4\ln\frac{\epsilon^2}{|z_{12}|^2} + \ln^2\frac{\epsilon^2}{|z_{12}|^2}\right)\right]\,.
    \end{split}
\end{equation}
We note that there is an additional contribution to the three-loop term of order $\nicefrac{c_G}{k}$~\cite{Georgiou:2016iom}
\begin{equation}
  \langle J^{a_1}(z_1)\Bar{J}^{a_2}(\bz_2)\rangle^{(3)}_\text{sub}=\frac{2\lambda^3c_G\delta_{a_1a_2}}{k}
  \left(\frac{1}{|z_{12}|^2}+\pi\delta^{(2)}(z_{12})\left(1-\ln\frac{\epsilon^2}{|z_{12}|^2}\right)\right)\,.
\end{equation}
However, this term is subleading when $c_G\gg k$.

\section{Anomalous Dimensions and Beta Functions}
\label{Section:Anomalous.Beta}

The objective of this section is to compute the leading $\nicefrac{c_G}{k}$ contribution  to the $\beta$-function and the anomalous dimensions of the current and composite operators. We will first evaluate these quantities using the results for the current operators from section~\ref{Section.currents}, and subsequently for the composite  operator from section~\ref{Section.Composite}.

\subsection{From the current operators}

At first we recap the following 3-loop result:
\begin{eqnarray}
  \langle J^{a_1}(z_1)J^{a_2}(z_2)\rangle^{(3)} &&= 
\frac{\lambda^3c_G^2\delta_{a_1a_2}}{z_{12}^2k^2}\left(-\frac{1}{2}\ln^2\frac{\epsilon^2}{|z_{12}|^2}+\ln\frac{\epsilon^2}{|z_{12}|^2}\right)\,,\\
    \langle J^{a_1}(z_1)\Bar{J}^{a_2}(\bz_2)\rangle^{(3)} &&= \frac{\lambda^3c_G^2\delta_{a_1a_2}}{k^2|z_{12}|^2}\left[\left(\ln\frac{\epsilon^2}{|z_{12}|^2}-2\right)\right.\nonumber\\
    &&\left.-\frac{\pi}{4}\delta^{(2)}(z_{12})\left(6-4\ln\frac{\epsilon^2}{|z_{12}|^2} + \ln^2\frac{\epsilon^2}{|z_{12}|^2}\right)\right]\,.
\end{eqnarray}
For the 2-loop case, only $\nicefrac{c_G}{k}$ terms appear~\cite{Georgiou:2015nka,Georgiou:2016iom}, which were restated in Eqs.\eqref{JJ.twoloop}, \eqref{JJbar.twoloop}.

 In fact, for $c_G\gg k$, an analog power counting like we did for the OO correlator in section \ref{Section.Composite.tHooft}, showing that the leading order 
 correction in $\nicefrac{k}{c_G}$ in the 't Hooft limit can be written in the form
\begin{equation}
    \frac{k}{c_G}\tilde{\lambda}^Lf_L(z_{12},\epsilon)\,.
\end{equation}
For the moment we only consider up to three loops. For the $JJ$ correlator, there is no 1-loop contribution, and we have
\begin{equation}
    \langle J^{a_1}(z_1)J^{a_2}(z_2)\rangle = \frac{\delta_{a_1a_2}}{z_{12}^2} \left[1+\frac{k}{c_G}\left(\Tilde{\lambda}^2\ln\frac{\epsilon^2}{|z_{12}|^2} + \Tilde{\lambda}^3\ln\frac{\epsilon^2}{|z_{12}|^2} - \frac{1}{2}\Tilde{\lambda}^3\ln^2\frac{\epsilon^2}{|z_{12}|^2} \right)\right]+\cdots\,.
\end{equation}
For the $J\bar J$ correlator we focus on the contact term, which reads
\begin{equation}
\begin{split}
    \langle J^{a_1}(z_1)\Bar{J}^{a_2}(\bz_2)\rangle &= -\pi\delta_{a_1a_2}\delta^{(2)}(z_{12})\frac{k}{c_G}\Tilde{\lambda}\left[1+\Tilde{\lambda}\left(1-\frac{1}{2}\ln\frac{\epsilon^2}{|z_{12}|^2}\right)\right.\\
    &\qquad\qquad\left.+\Tilde{\lambda}^2\left(\frac{3}{2}-\ln\frac{\epsilon^2}{|z_{12}|^2} + \frac{1}{4}\ln^2\frac{\epsilon^2}{|z_{12}|^2}\right)\right]+\cdots\,.
 \end{split}
\end{equation}
Next we perform the renormalization, where the bare and renormalized quantities are related as usual by
\begin{equation}
    J_0^a = Z^{\nicefrac{1}{2}}J^a\,,\,\Bar{J}_0^a = Z^{\nicefrac{1}{2}}\Bar{J}^a\,,\,\Tilde{\lambda}_0 = Z_1\Tilde{\lambda}\,.
\end{equation}
We make the following ansatz
\begin{eqnarray}
    Z^{-1} &=& 1 - \frac{k}{c_G}\left(c_1\Tilde{\lambda}^2+c_2\Tilde{\lambda}^3\right)\ln(\epsilon^2\mu^2)-\frac{k}{c_G}d_1\Tilde{\lambda}^3\ln^2(\epsilon^2\mu^2)\,,\\
    Z_1 &=& 1 - (c_3\Tilde{\lambda}+c_4\Tilde{\lambda}^2)\ln(\epsilon^2\mu^2)-d_2\tilde{\lambda}^2\ln^2(\epsilon^2\mu^2)\,.
\end{eqnarray}
Now we require the renormalized correlators
\begin{equation}
    \begin{split}
    &\langle J^{a_1}(z_1)J^{a_2}(z_2)\rangle =  \langle J_0^{a_1}(z_1)J_0^{a_2}(z_2) \rangle Z^{-1} \,,\\
    &\langle J^{a_1}(z_1)\Bar{J}^{a_2}(\bar z_2)\rangle =  \langle J_0^{a_1}(z_1)\Bar{J}_0^{a_2}(\bar z_2)\rangle Z^{-1} \,,
    \end{split}
\end{equation}
to be independent of the cutoff $\epsilon$. To proceed we keep only terms of order $\nicefrac{k}{c_G}$ and up to $\ln^2$ and $\Tilde{\lambda}^3$
\begin{eqnarray}
    &&\langle J^{a_1}(z_1)J^{a_2}(z_2)\rangle= \langle J_0^{a_1}(z_1)J_0^{a_2}(z_2)\rangle Z^{-1} \nonumber\\
    &=& \frac{\delta_{a_1a_2}}{z_{12}^2}\left[1+\frac{k}{c_G}\left(\Tilde{\lambda}_0^2\ln\frac{\epsilon^2}{|z_{12}|^2} + \Tilde{\lambda}_0^3\ln\frac{\epsilon^2}{|z_{12}|^2}-\frac{1}{2}\Tilde{\lambda}_0^3\ln^2\frac{\epsilon^2}{|z_{12}|^2}\right)\right]\nonumber\\
    &&\times\left[1 - \frac{k}{c_G}\left(c_1\Tilde{\lambda}^2+c_2\Tilde{\lambda}^3\right)\ln(\epsilon^2\mu^2)-\frac{k}{c_G}d_1\Tilde{\lambda}^3\ln^2(\epsilon^2\mu^2)\right]
    \nonumber\\
    &=& \frac{\delta_{a_1a_2}}{z_{12}^2} \left[1+\frac{k}{c_G}\left(  -c_1\Tilde{\lambda}^2\ln(\epsilon^2\mu^2)-c_2\Tilde{\lambda}^3\ln(\epsilon^2\mu^2)+\Tilde{\lambda}^2\ln\frac{\epsilon^2}{|z_{12}|^2} + \Tilde{\lambda}^3\ln\frac{\epsilon^2}{|z_{12}|^2}\right.\right.\nonumber\\
    &&\qquad\left.\left.-2 c_3\tilde{\lambda}^3\ln(\epsilon^2\mu^2)\ln\frac{\epsilon^2}{|z_{12}|^2}-d_1\tilde{\lambda}^3\ln^2(\epsilon^2\mu^2)  -\frac{1}{2}\tilde{\lambda}^3\ln^2\frac{\epsilon^2}{|z_{12}|^2}+...\right)\right]\,.
\end{eqnarray}
We can find 
\begin{equation}
\label{Renorm.JJ}
c_1 = 1\,,\quad c_2=1\,,\quad c_3=-\frac{1}{2}\,,\quad d_1=\frac{1}{2}\,.
\end{equation}
This tells us that the anomalous dimension is
\begin{equation}
\label{anomalous.J}
    \gamma^{(J)} = \frac{d\ln Z^{\frac{1}{2}}}{d\ln\mu} = \frac{k}{c_G}\left(\Tilde{\lambda}^2+\Tilde{\lambda}^3+O(\Tilde{\lambda}^4)\right)\,.
\end{equation}
Similarly for $J\bar J$, we keep only terms of order $\nicefrac{k}{c_G}$ and up to $\ln^2$ and $\Tilde{\lambda}^3$
\begin{eqnarray}
    &&\langle J^{a_1}(z_1)\Bar{J}^{a_2}(\bz_2)\rangle= \langle J_0^{a_1}(z_1)\Bar{J}_0^{a_2}(\bz_2)\rangle Z^{-1} \nonumber\\
   &&= -\pi\delta_{a_1a_2}\delta^{(2)}(z_{12})\frac{k}{c_G}\Tilde{\lambda}_0\left[1+\Tilde{\lambda}_0\left(1-\frac{1}{2}\ln\frac{\epsilon^2}{|z_{12}|^2}\right)+\Tilde{\lambda}_0^2\left(\frac{3}{2}-\ln\frac{\epsilon^2}{|z_{12}|^2}+\frac{1}{4}\ln^2\frac{\epsilon^2}{|z_{12}|^2} \right)\right]\nonumber\\
    &&\qquad\times\left[1 - \frac{k}{c_G}\left(\Tilde{\lambda}^2+\Tilde{\lambda}^3\right)\ln(\epsilon^2\mu^2)-\frac{k}{c_G}d_1\tilde{\lambda}^3\ln^2(\epsilon^2\mu^2)\right]\nonumber\\
    &&=-\pi\delta_{a_1a_2}\delta^{(2)}(z_{12})\frac{k}{c_G}\Tilde{\lambda}\left(1 - (c_3\Tilde{\lambda}+c_4\Tilde{\lambda}^2)
    \ln(\epsilon^2\mu^2)-d_2\tilde{\lambda}^2\ln^2(\epsilon^2\mu^2)\right)\nonumber\\
    &&\times\left[1+\Tilde{\lambda}-c_3\Tilde{\lambda}^2\ln(\epsilon^2\mu^2)-\frac{1}{2}\Tilde{\lambda}\ln\frac{\epsilon^2}{|z_{12}|^2}+\frac{1}{2}c_3\tilde{\lambda}^2\ln(\epsilon^2\mu^2)\ln\frac{\epsilon^2}{|z_{12}|^2} + \Tilde{\lambda}^2\left(\frac{3}{2}-\ln\frac{\epsilon^2}{|z_{12}|^2}+\frac{1}{4}\ln^2\frac{\epsilon^2}{|z_{12}|^2} \right)\right]\nonumber\\
    &&= -\pi\delta_{a_1a_2}\delta^{(2)}(z_{12})\frac{k}{c_G}\Tilde{\lambda} \left[1+\Tilde{\lambda}+\frac{3}{2}\Tilde{\lambda}^2 - \Tilde{\lambda}\left(c_3\ln(\epsilon^2\mu^2)+\frac{1}{2}\ln\frac{\epsilon^2}{|z_{12}|^2}\right)- \Tilde{\lambda}^2(2c_3+c_4)\ln(\epsilon^2\mu^2)\right.\nonumber\\
    &&\qquad\qquad\qquad\qquad\left.-\Tilde{\lambda}^2\ln\frac{\epsilon^2}{|z_{12}|^2}-d_2\tilde{\lambda}^2\ln^2(\epsilon^2\mu^2)+c_3\tilde{\lambda}^2\ln(\epsilon^2\mu^2)\ln\frac{\epsilon^2}{|z_{12}|^2} + \frac{1}{4}\tilde{\lambda}^2\ln^2\frac{\epsilon^2}{|z_{12}|^2}\right]\,.
\end{eqnarray}
For this to be independent of the cutoff $\epsilon$ we find
\begin{equation}
    c_3 = -\frac{1}{2}\,,\quad c_4 = 0\,,\quad d_2 = -\frac{1}{4}\,,
\end{equation}
where we note that the value of $c_3$ is consistent with the one in \eqref{Renorm.JJ}.
Therefore the beta function in the 't Hooft limit is\footnote{Note that to order $\Tilde\lambda^3$ there is an additional contribution of the order $\nicefrac{k}{c_G}$, see Eq.(3.14) in \cite{Georgiou:2016iom}
\begin{equation*}
  \beta^{\Tilde{\lambda}} = -\frac{\Tilde{\lambda}^2}{2}+\frac{k}{c_G}\Tilde\lambda^3+O(\Tilde{\lambda}^4)\,,
\end{equation*}
with no additional perturbative fixed point - apart from the UV fixed point at $\Tilde\lambda=0$.} 
\begin{equation}
    \beta^{\Tilde{\lambda}} =
    \frac{d\Tilde\lambda}{d\ln\mu^2}
    = -\frac{1}{2}\Tilde{\lambda}\frac{d\ln Z_1}{d\ln\mu} = -\frac{\Tilde{\lambda}^2}{2}+O(\Tilde{\lambda}^4) \,.\label{betafromJJ}
\end{equation}
In brief, to leading order in the 't Hooft limit the $\beta$-function is quadratic exact (at least to cubic order) and the anomalous dimension of the current operator vanishes at leading order;~\eqref{anomalous.J} is order $\nicefrac{k}{c_G}$.

\subsection{From the composite operator}

We can also use the OO correlators to determine the beta function. In this case we find that our 3-loop computation can fix the term at order $\tilde{\lambda}^4$ at leading order in $\nicefrac{c_G}{k}$. The OO correlator up to $\tilde{\lambda}^3$ at leading order in $\nicefrac{c_G}{k}$ is given by
\begin{equation}
    \langle O(z_1)O(z_2) \rangle = \frac{\dim G}{|z_{12}|^4}\left(1-2\tilde{\lambda}\ln\frac{\epsilon^2}{|z_{12}|^2} + \frac{5}{2}\tilde{\lambda}^2\ln^2\frac{\epsilon^2}{|z_{12}|^2} - \frac{5}{2}\tilde{\lambda}^3\ln^3\frac{\epsilon^2}{|z_{12}|^2}+\frac{10\zeta(3)\tilde\lambda^3}{3}
    \right)\,.
\end{equation}
The anomalous dimension will be given by
\begin{equation}
\label{anomalous.O.first}
    \gamma_O = -2\tilde{\lambda} + O(\tilde{\lambda}^4)
\end{equation}
hence it is linearly exact (at least to cubic order).
Moreover, we know from \cite{Kutasov:1989dt}
\begin{equation}
    \gamma_O = 2\frac{d\beta^{\tilde{\lambda}}}{d\tilde{\lambda}} + \beta^{\tilde{\lambda}}\frac{d\ln \cal G}{d\tilde{\lambda}}\,.
\end{equation}
We also notice that the finite part of the Zamolodchikov metric leads to
\begin{equation}
\label{betafromOO}
    \beta^{\tilde{\lambda}} = -\frac{\tilde{\lambda}^2}{2} + O(\tilde{\lambda}^5)\,,
\end{equation}
so the above result extends \eqref{betafromJJ} to quartic order.

This result parallels the findings of~\cite{Ludwig:2002fu,Bykov:2022hvt}, which examine the chiral Gross--Neveu (non-Abelian Thirring) model at $k=0$, effectively corresponding to the $c_G \to \infty$ limit. In those works, the quintic-order was found to be non-vanishing but subleading in the 't Hooft limit we study.\footnote{Compared with the one-loop contribution, it is of order $\nicefrac{1}{N^2}$ for $G=\mathrm{SU}(N)$ and $\nicefrac{1}{N}$ for $G=\mathrm{SO}(N)\text{ and }\mathrm{SP}(N)$.}

\section{Concluding and Outlook}
\label{Section:Concl.Out}

In the present work, we considered the non-Abelian bosonized Thirring model at large $N$, which is characterized by the coupling constant $\lambda$, the integer level $k$ of the WZW model, and the semi-simple group $G$. Our goal was to search for non-trivial fixed points of the order of $\nicefrac1N$, following~\cite{Dashen:1974gu}, for the non-Abelian fermionic Thirring model. In this regime the all-loop effective action of the non-Abelian bosonized Thirring model found in~\cite{Georgiou:2016iom,Sfetsos:2013wia}, valid in the semi-classical regime $\nicefrac{c_G}{k}\ll1$, is inapplicable.

Using conformal perturbation theory, we computed the two-point correlation functions of the currents and of the driving operators, which enabled us to derive their anomalous dimensions and the $\beta$-function. For large $N$ and in the 't Hooft limit, the resulting expressions simplified drastically. In particular: 
\begin{itemize}
\item
The anomalous dimensions of the current and the composite operator were given by \eqref{anomalous.J} and \eqref{anomalous.O.first}, respectively.
\item
The $\beta$-function is quadratic exact to quartic order~\eqref{betafromOO}. Hence, to this order there is no additional $\nicefrac1N$ fixed point, contrary to the claim of Dashen--Frishman in~\cite{Dashen:1974gu} and in alignment with the findings of Destri \& de Vega of~\cite{Destri,Destri.Vega} for the Fermionic non-Abelian Thirring model.
\item
A comment is in order regarding the points $\lambda=\pm1$, which appear in the semi-classical regime $\nicefrac{c_G}{k}\ll1$. In this limit, they act as fixed points of the symmetry \eqref{symmetry.general} and correspond to the non-Abelian T-dual and pseudo-dual models~[\citenum{Sfetsos:2013wia}, \citenum{Georgiou:2016iom}]. However, our perturbative expansion is valid in the double scaling limit of \eqref{lambda.tHooft}, where the points $\lambda=\pm1$ cannot be probed.
\end{itemize}

A natural direction to extend our work is to prove that for $c_G\gg1$, the exact expressions for the anomalous dimensions as well as the $\beta$-function truncate to \eqref{anomalous.J}, \eqref{anomalous.O.first} and \eqref{betafromOO}. An analogue property was shown in~\cite{Gross:1974jv} and~\cite{Mitter:1973dwt} for the Gross--Neveu and the massive Thirring model respectively.

Another extension would be to compute three-point functions of current operators at large $c_G$ and deduce the OPEs as well as the equal time commutators of the currents. This will enable us to derive the underlying symmetry algebra at large $N$ as it was done in~\cite{Georgiou:2016iom} for large $k$ and previous in~\cite{Ashok:2009xx,Konechny:2010nq} for supercosets.

Finally, another potential extension includes the non-Abelian Thirring model for unequal levels $k_{L,R}$ of the left- and right-current algebras. For $k_{L,R}\gg1$, its all-loop renormalization has been worked out in~\cite{LeClair:2001yp} (see also~\cite{Georgiou:2016zyo}), while its effective action was constructed in~\cite{Georgiou:2017jfi} and proved in~\cite{Georgiou:2016zyo,Sagkrioti:2018rwg}. In contrast to the equal-level case, which has a Fermionic analogue at level one, the model interpolates between two exact 
CFTs as suggested in~\cite{LeClair:2001yp,chiraliquids} and showed in~\cite{Georgiou:2017jfi}. It would be interesting to study this class of  asymmetric deformations for large values of $c_G$, unveiling potential additional fixed points as well as the symmetry algebra.

\acknowledgments

The authors would like to thank Konstantinos Sfetsos for useful discussions. The work of SL was supported by  the Natural Science Foundation of China (Grant No. 12522504). We acknowledge the use of Anthropic Claude (Opus 5.5) as a research aid to help identify errors.

\appendix

\section{An integral toolkit}
\label{appendix:A}

In this appendix we compute in detail the basic integrals in the computations of the loop-corrections of correlation functions. 
For reader's convenience the final expressions will be framed.

An important tool we are going to use is the Green's theorem written in complex coordinates:
\begin{equation}
\label{Greenthm}
\int_S d^2 z (\partial_{z} A+\partial_{\bz} B) = \frac{i}{2} \oint_{\partial S} (Ad\bz-Bdz)\,,
\end{equation}
where $z=x+i y$ and $d^2z=dx dy$. In addition we will need the following identity, found using Green's theorem \eqref{Greenthm}
\begin{equation}
\partial_z\left(\frac{1}{\bz}\right)=\partial_\bz\left(\frac{1}{z}\right)=\pi\delta^{(2)}(z)\,,\quad \delta^{(2)}(z)=\delta(x)\delta(y)\,,
\end{equation}
as well as descendants of it, for example
\begin{equation}
\partial_\bz\left(\frac{1}{z^2}\right)=-\pi\partial_z\delta^{(2)}(z)\,,\quad \partial_z\left(\frac{1}{\bz^2}\right)=-\pi\partial_\bz\delta^{(2)}(z)\,.
\end{equation}

The basic integral is given by
\begin{equation}
\label{basic.int}
\boxed{
\int\frac{d^2z}{(z-z_1)(\bar z_2-\bar z)}=\pi\ln\frac{|z_{12}|^2}{R^2}\,,}
\end{equation}
where $\mathcal{D}$ is a disc of radius $R\gg|z_{1,2}|$. The integral may be found using Green's theorem~\eqref{Greenthm} with
\begin{equation}
A=\frac{1}{\bar z-\bar z_2}\ln\frac{|z-z_1|^2}{R^2}\,,\quad B=0\,,
\end{equation}
where the arc contribution $|z|=R$ turns out to vanish. Descendants of the above integral include
\begin{equation}
\boxed{
\int\frac{d^2z}{(z-z_1)^2(\bar z-\bar z_2)}=-\frac{\pi}{z_{12}}\,,\quad \int\frac{d^2z}{(z-z_1)(\bar z-\bar z_2)^2}=\frac{\pi}{\bar z_{12}}\,.}
\end{equation}

An important point are the UV-divergences, these are treated by using the $\epsilon$ cut-off when the two external points coincide. In \eqref{basic.int}, when the external points coincide we obtain
\begin{equation}
\label{basic.int.epsilon}
\boxed{
\int\frac{d^2z}{(z-z_1)(\bar z_1-\bar z)}=\pi\ln\frac{\epsilon^2}{R^2}\,.}
\end{equation}

\subsection*{Another basic integral}
We are also interested in computing the following integral  
\begin{equation}
\label{Integral123}
I(z_1, z_2,z_3)=\int\frac{d^2z}{(z-z_1)(\bar z-\bar z_2)}\ln\frac{|z-z_3|^2}{R^2}\,,
\end{equation}
where $\mathcal{D}$ is a disc of radius $R\gg|z_{1,2,3}|$ and by construction it satisfies the property
\begin{equation}
\label{Integral123.real}
I(z_1,z_2,z_3)=I^*(z_2,z_1,z_3)\,.
\end{equation}
As we will prove, this it is given by
\begin{equation}
\label{Integral123.final}
\boxed{
I(z_1,z_2,z_3)=-\frac{\pi}{2}\left(\ln\frac{|z_{12}|^2}{R^2}\ln\frac{|z_{23}|^2}{R^2}+\ln\frac{|z_{12}|^2}{R^2}\ln\frac{|z_{13}|^2}{R^2}
-\ln\frac{|z_{13}|^2}{R^2}\ln\frac{|z_{23}|^2}{R^2}\right)-2\pi i D_2(z)\,,}
\end{equation}
where $D_2(z)$ is the Bloch--Wigner function\footnote{\label{Bloch}The Bloch--Wigner function is real and single-valued, satisfying several properties 
\begin{equation*}
D_2(z)=-D_2(1-z)=
-D_2\left(\frac1z\right)=-D_2\left(-\frac{z}{1-z}\right)\,.
\end{equation*}
}
\begin{equation}
D_2(z)=\Im\left(\text{Li}_2(z)\right)+\arg(1-z)\ln|z|\,,
\end{equation}
where $\displaystyle z=\frac{z_{13}}{z_{12}}$ and $\text{Li}_2(z)$ is the polylogarithm of order-2 defined by
\begin{equation}
\text{Li}_2(z)=-\int_0^z dt\frac{\ln(1-t)}{t}\,.
\end{equation}

\paragraph{Proof:} At first we consider Green's theorem \eqref{Greenthm} with
\begin{equation}
A=\frac{\ln\frac{|z-z_1|^2}{R^2}\ln\frac{|z-z_3|^2}{R^2}}{\bar z-\bar z_2}\,,\quad B=0\,,
\end{equation}
yielding 
\begin{equation}
\partial A=\frac{\ln\frac{|z-z_3|^2}{R^2}}{(z-z_1)(\bar z-\bar z_2)}+\frac{\ln\frac{|z-z_1|^2}{R^2}}{(z-z_3)(\bar z-\bar z_2)}
+\pi\delta^{(2)}(z-z_2)\ln\frac{|z-z_1|^2}{R^2}\ln\frac{|z-z_3|^2}{R^2}\,.
\end{equation}
Using \eqref{Greenthm} we find
\begin{equation}
\label{Integral123.prop}
I(z_1, z_2,z_3)+I(z_3, z_2,z_1)=-\pi\ln\frac{|z_{12}|^2}{R^2}\ln\frac{|z_{23}|^2}{R^2}\,,
\end{equation}
where it can be easily seen that the arc contribution $|z|=R$ vanishes. Alternatively we can reach \eqref{Integral123.prop} starting from the integral 
\begin{equation}
\frac{1}{\pi}\int \frac{d^2z_{4,5}}{z_{34}z_{15}\bar z_{25}\bar z_{45}}=
\int d^2z_5\frac{\ln\frac{|z_5-z_3|^2}{R^2}}{z_{15}\bar z_{25}}=I(z_1,z_2,z_3)
\end{equation}
and integrating with respect to $z_4$, where in the first equality we used \eqref{basic.int}. Equivalently we can rewrite it as
\begin{equation}
\begin{split}
I(z_1,z_2,z_3)&=\frac{1}{\pi}\int\frac{d^2z_{4,5}}{z_{34}z_{15}}\frac{1}{\bar z_{24}}\left(\frac{1}{\bar z_{45}}-\frac{1}{\bar z_{25}}\right)\\
&=-\pi\ln\frac{|z_{12}|^2}{R^2}\ln\frac{|z_{23}|^2}{R^2}-I(z_3,z_2,z_1)\,,
\end{split}
\end{equation}
where in the second equality we integrated with respect to $z_5$.

Starting from \eqref{Integral123.prop}, 
cycling permuting $z_{1,2,3}$ and using \eqref{Integral123.real}, we easily find its real part 
\begin{equation}
\label{Integral123.Re}
\Re(I(z_1, z_2,z_3))=-\frac{\pi}{2}\left(\ln\frac{|z_{12}|^2}{R^2}\ln\frac{|z_{23}|^2}{R^2}+\ln\frac{|z_{12}|^2}{R^2}\ln\frac{|z_{13}|^2}{R^2}
-\ln\frac{|z_{13}|^2}{R^2}\ln\frac{|z_{23}|^2}{R^2}\right)
\end{equation}
and that the imaginary part of $I(z_1, z_2,z_3)$ is totally antisymmetric in $z_{1,2,3}$. 

To compute the imaginary part  we differentiate \eqref{Integral123} with respect to $z_3$ and $\bar z_3$ yielding
\begin{equation}
\begin{split}
&\partial_{z_3}I(z_1,z_2,z_3)=-\int\frac{d^2z}{(z-z_1)(z-z_3)(\bar z-\bar z_2)}=\frac{\pi}{z_{13}}\ln\frac{|z_{12}|^2}{|z_{23}|^2}\,,\\ 
&\partial_{\bar z_3}I(z_1,z_2,z_3)=-\int\frac{d^2z}{(z-z_1)(\bar z-\bar z_2)(\bar z-\bar z_3)}=\frac{\pi}{\bar z_{23}}\ln\frac{|z_{12}|^2}{|z_{13}|^2}\,.
\end{split}
\end{equation}
From the above and \eqref{Integral123.Re} we find
\begin{equation}
\begin{split}
&i\partial_{z_3}\Im(I(z_1,z_2,z_3))=\frac\pi2\left(\frac{1}{z_{13}}\ln\frac{|z_{12}|^2}{|z_{23}|^2}-
\frac{1}{z_{23}}\ln\frac{|z_{12}|^2}{|z_{13}|^2}\right)\,,\\
&i\partial_{\bar z_3}\Im(I(z_1,z_2,z_3))=\frac\pi2\left(\frac{1}{\bar z_{23}}\ln\frac{|z_{12}|^2}{|z_{13}|^2}-
\frac{1}{\bar z_{13}}\ln\frac{|z_{12}|^2}{|z_{23}|^2}\right)\,.
\end{split}
\end{equation}
The above system is integrable and upon integrating it we find\footnote{In principle, the imaginary part could also include an additional unknown function of $z_1$ and $z_2$; however, as it turns out, this function vanishes. To prove this, we set $z_3$ equal to either $z_1$ or $z_2$, where the result of the corresponding integral is known to be real and is given by Eq.~\eqref{spin-off.I}. Comparing this with \eqref{Integral123.final}, we can readily check that the real part suffices; hence, the arbitrary function in the imaginary part vanishes.}
\begin{equation}
\label{Integral123.Im}
\Im(I(z_1,z_2,z_3))=-2\pi D_2(z)\,,\quad z=\frac{z_{13}}{z_{12}}\,.
\end{equation} 
Finally, combining the above and \eqref{Integral123.Re}, we reach \eqref{Integral123.final}.

\subsection*{Properties of \eqref{Integral123.final}}
$\bullet$ Using the Bloch--Wigner function properties, given in footnote \ref{Bloch}, one can show that  \eqref{Integral123.real}, \eqref{Integral123.prop} are satisfied and that the imaginary part \eqref{Integral123.Im} is totally antisymmetric as anticipated.

\noindent
$\bullet$ Using \eqref{Integral123.final} and footnote \ref{Bloch} we can prove the identity
\begin{equation}
\label{Integral123.prop1}
I(z_1, z_2,z_3)+I(z_1, z_3,z_2)=-\pi\ln\frac{|z_{12}|^2}{R^2}\ln\frac{|z_{13}|^2}{R^2}\,,
\end{equation}
$\bullet$ From the integral \eqref{Integral123.final} we easily find
\begin{equation}
\label{spin-off.I}
I(z_1,z_2,z_2)=-\frac\pi2\ln^2\frac{|z_{12}|^2}{R^2}\,,\quad 
I(z_1,z_2,z_1)=-\frac\pi2\ln^2\frac{|z_{12}|^2}{R^2}\,,
\end{equation}
in agreement with Eq. (2.35) of \cite{Georgiou:2016iom}. Also when $z_{1,2}$ coincide in \eqref{Integral123.final}, we use the $\epsilon$ cut-off as in \eqref{basic.int.epsilon}
\begin{equation}
\label{Iwith2equal3}
    I(z_1,z_1,z_3) = -\pi\ln\frac{\epsilon^2}{R^2}\ln\frac{|z_{13}|^2}{R^2} + \frac{\pi}{2}\ln^2\frac{|z_{13}|^2}{R^2}\,.
\end{equation}
One can also derive $I(z_1, z_2, z_3)$ with respect to $z_1$ and $\bar z_2$ and find for $z_{1,2,3}$ being distinct points
\begin{equation}
\boxed{
\begin{split}
    \int\frac{d^2z}{(z-z_1)^2(\bar z-\bar z_2)}\ln\frac{|z-z_3|^2}{R^2} &= \partial_{z_1} I(z_1, z_2, z_3)\\
    &= -\frac{\pi}{z_{12}}\ln\frac{|z_{23}|^2}{R^2} - \frac{\pi}{z_{13}}\ln\frac{|z_{12}|^2}{|z_{23}|^2}\,,\\
    \int\frac{d^2z}{(z-z_1)(\bar z-\bar z_2)^2}\ln\frac{|z-z_3|^2}{R^2} &= \partial_{\bar z_2}I(z_1, z_2, z_3)\\
    &= \frac{\pi}{\bar z_{12}}\ln\frac{|z_{13}|^2}{R^2} - \frac{\pi}{\bar z_{23}}\ln\frac{|z_{12}|^2}{|z_{13}|^2}\,.
    \end{split}}
\end{equation}
From the above we find
\begin{equation}
\boxed{
 \int\frac{d^2z}{(z-z_1)^2(\bar z-\bar z_2)^2}\ln\frac{|z-z_3|^2}{R^2}=\pi^2\delta^{(2)}(z_{12})\ln\frac{|z_{13}|^2}{R^2}
 -\frac{\pi}{z_{12}\bar z_{23}}+\frac{\pi\bar z_{13}}{z_{13}\bar z_{12}\bar z_{23}}\,,}
\end{equation}
in agreement with (C.13) of \cite{Georgiou:2019jcf}.

The cases where two of the three points coincide will be computed in a slightly different way. First we recall \eqref{spin-off.I}
\begin{equation}
\boxed{
  I(z_1,z_2,z_2) =  \int \frac{ d^2z}{(z-z_1)(\bar z-\bar z_2)}\ln\frac{|z-z_2|^2}{R^2} =  -\frac{\pi}{2}\ln^2\frac{|z_{12}|^2}{R^2}\,,}
\end{equation}
and by deriving with respect to $z_1$ and $\bar z_2$ we find
\begin{equation}
\boxed{
\begin{split}
    \int\frac{d^2z}{(z-z_1)^2(\bar z-\bar z_2)}\ln\frac{|z-z_2|^2}{R^2} &= \partial_{z_1} I(z_1,z_2,z_2)\\
    &= -\frac{\pi}{z_{12}}\ln\frac{|z_{12}|^2}{R^2}\,,\\
    \int\frac{d^2z}{(z-z_1)(\bar z-\bar z_2)^2}\ln\frac{|z-z_2|^2}{R^2} &= \partial_{\bar z_2} \left(I(z_1,z_2,z_2) + \int\frac{d^2z}{(z-z_1)(\bar z-\bar z_2)}\right)\\
    &= \frac{\pi}{\bar z_{12}}\left(1+\ln\frac{|z_{12}|^2}{R^2}\right)\,.
    \end{split}}
\end{equation}

From the latter we can easily find
\begin{equation}
\label{log.fraction}
\boxed{
\int\frac{d^2z}{(z-z_1)^2(\bar z-\bar z_2)^2}\ln\frac{|z-z_2|^2}{R^2}=\pi^2\delta^{(2)}(z_{12})\left(1+\ln\frac{|z_{12}|^2}{R^2}\right)+
\frac{\pi}{|z_{12}|^2}\,,}
\end{equation}
which agrees with (B.12) of \cite{Georgiou:2016iom}\,.

Now if we consider Green's theorem~\eqref{Greenthm} with
\begin{equation}
A=\frac{1}{(\bar z-\bar z_1)^2}\ln\frac{|z-z_1|^2}{R^2}\ln\frac{|z-z_2|^2}{R^2}\,,\quad B=0\,,
\end{equation}
we find
\begin{equation}
\begin{split}
    &\int\frac{d^2z}{(z-z_1)(\bar z-\bar z_1)^2}\ln\frac{|z-z_2|^2}{R^2} + \int\frac{d^2z}{(z-z_2)(\bar z-\bar z_1)^2}\ln\frac{|z-z_1|^2}{R^2}\\
    &+ \int d^2z\,\partial_{z}\left(\frac{1}{(\bz-\bar z_1)^2}\right)\ln\frac{|z-z_1|^2}{R^2}\ln\frac{|z-z_2|^2}{R^2} = 0\,.
 \end{split}
\end{equation}
We now have
\begin{equation}
\begin{split}
    &\int\frac{d^2z}{(z-z_1)(\bar z-\bar z_1)^2}\ln\frac{|z-z_2|^2}{R^2} - \frac{\pi}{\bar z_{12}}\left(1+\ln\frac{|z_{12}|^2}{R^2}\right)\\
    &= \pi\int d^2 z \partial_{\bz}(\delta^{(2)}(z-z_1))\ln\frac{|z-z_1|^2}{R^2}\ln\frac{|z-z_2|^2}{R^2}\\
    &= -\pi\int d^2 z \delta^{(2)}(z-z_1)\partial_{\bz}\left(\ln\frac{|z-z_1|^2}{R^2}\ln\frac{|z-z_2|^2}{R^2}\right)\\
    &= -\frac{\pi}{\bar z_{12}}\ln\frac{\epsilon^2}{R^2}-\pi\int d^2z \frac{\delta^{(2)}(z-z_1)}{\bar z-\bar z_1}\ln\frac{|z-z_2|^2}{R^2}\,.
 \end{split}
\end{equation}
The second term in the last line can be written as
\begin{equation}
    -\pi\ln\frac{|z_{12}|^2}{R^2}\int d^2z \frac{\delta^{(2)}(z-z_1)}{\bar z-\bar z_1}\,,
\end{equation}
which is vanishing. Alternatively, it can be viewed as
\begin{equation}
    0 = \partial_{\bar z_1}\int \frac{d^2z}{(z-z_1)(\bz-\bar z_1)} = -\pi\int d^2z  \frac{\delta^{(2)}(z-z_1)}{ \bar z-\bar z_1} + 
    \int \frac{d^2z}{(z-z_1)(\bz-\bar z_1)^2}\,,
\end{equation}
since the last integral is vanishing, after an angular integration.
So we find
\begin{equation}
\boxed{
    \int\frac{d^2z}{(z-z_1)(\bar z-\bar z_1)^2}\ln\frac{|z-z_2|^2}{R^2} = \frac{\pi}{\bar z_{12}}\left(1-\ln\frac{\epsilon^2}{|z_{12}|^2}\right)\,.}
\end{equation}

Now we consider the following integral:
\begin{equation}
    \int \frac{d^2z}{(z-z_1)(\bz-\bar z_2)}\ln^2\frac{|z-z_1|^2}{R^2}\,.
\end{equation}
One can consider Green's theorem~\eqref{Greenthm} with
\begin{equation}
A=\frac{1}{\bz-\bar z_2}\ln^3\frac{|z-z_1|^2}{R^2}\,,\quad B=0\,,
\end{equation}
and find
\begin{equation}
    \int \frac{d^2z}{(z-z_1)(\bz-\bar z_2)}\ln^2\frac{|z-z_1|^2}{R^2} = -\frac{\pi}{3}\ln^3\frac{|z_{12}|^2}{R^2}\,.
\end{equation}
Deriving the latter with respect to $z_1$ and $\bar z_2$ and using \eqref{log.fraction} we find
\begin{equation}
\boxed{
\begin{split}
    &\int \frac{d^2z}{(z-z_1)^2(\bz-\bar z_2)^2}\ln^2\frac{|z-z_1|^2}{R^2}\\
    &= \frac{2\pi}{|z_{12}|^2}\left(1+\ln\frac{|z_{12}|^2}{R^2}\right) + 2\pi^2\delta^{(2)}(z_{12})\left(1+\ln\frac{|z_{12}|^2}{R^2} + \frac{1}{2}\ln^2\frac{|z_{12}|^2}{R^2}\right)\,.
    \end{split}}
\end{equation}

A slightly more difficult integral is
\begin{equation}
\label{Integral.zeta3}
    \int \frac{d^2z}{(z-z_1)(\bz-\bar z_1)}\ln^2\frac{|z-z_2|^2}{R^2}\,.
\end{equation}
Computing its $z_2$ and $\bar z_2$ derivatives respectively we find
\begin{equation}
\begin{split}
\label{derx2}
    &\partial_{z_2}\int \frac{d^2z}{(z-z_1)(\bz-\bar z_1)}\ln^2\frac{|z-z_2|^2}{R^2}\\
    &= -2\int \frac{d^2z}{(z-z_1)(z-z_2)(\bz-\bar z_1)}\ln\frac{|z-z_2|^2}{R^2}\\
    &= \frac{2}{z_{12}}\int d^2z \left(\frac{1}{(z-z_2)(\bz-\bar z_1)}-\frac{1}{(z-z_1)(\bz-\bar z_1)}\right)\ln\frac{|z-z_2|^2}{R^2}\\
    &= -\frac{2\pi}{z_{12}}\left(-\ln\frac{\epsilon^2}{R^2}\ln\frac{|z_{12}|^2}{R^2}+\ln^2\frac{|z_{12}|^2}{R^2}\right)=
    -\frac{2\pi}{z_{12}}\ln\frac{|z_{12}|^2}{R^2}\ln\frac{|z_{12}|^2}{\epsilon^2}\,,
 \end{split}
 \end{equation}
    and 
\begin{equation}
\begin{split}
    \label{derx2bar}
    &\partial_{\bar z_2}\int \frac{d^2z}{(z-z_1)(\bz-\bar z_1)}\ln^2\frac{|z-z_2|^2}{R^2}\\
        &= -2\int \frac{d^2z}{(z-z_1)(\bar z-\bar z_2)(\bz-\bar z_1)}\ln\frac{|z-z_2|^2}{R^2}\\
    &= -\frac{2\pi}{\bar z_{12}}\left(-\ln\frac{\epsilon^2}{R^2}\ln\frac{|z_{12}|^2}{R^2}+\ln^2\frac{|z_{12}|^2}{R^2}\right)
    =  -\frac{2\pi}{\bar z_{12}}\ln\frac{|z_{12}|^2}{R^2}\ln\frac{|z_{12}|^2}{\epsilon^2}\,.
 \end{split}
 \end{equation}
One can integrate \eqref{derx2}, \eqref{derx2bar} and, due to translational invariance, obtain the $z_{12}$-dependence of \eqref{Integral.zeta3} 
\begin{equation}
    \int \frac{d^2z}{(z-z_1)(\bz-\bar z_1)}\ln^2\frac{|z-z_2|^2}{R^2} = -\pi\ln\frac{\epsilon^2}{R^2}\ln^2\frac{|z_{12}|^2}{R^2}+\frac{2\pi}{3}\ln^3\frac{|z_{12}|^2}{R^2}+c\,,
\end{equation}
up to a constant $c$ that can be evaluated by a direct computation of \eqref{Integral.zeta3} using polar coordinates. The end result is given by
\begin{equation}
\label{logsquare}
\boxed{
    \int \frac{d^2z}{(z-z_1)(\bz-\bar z_1)}\ln^2\frac{|z-z_2|^2}{R^2} = -\pi\ln\frac{\epsilon^2}{R^2}\ln^2\frac{|z_{12}|^2}{R^2}+\frac{2\pi}{3}\ln^3\frac{|z_{12}|^2}{R^2}+
    4\pi\zeta(3)\,,}
\end{equation}
where $\zeta(s)=\sum\limits_{n\geqslant1}\frac{1}{n^s}$ is the Riemann zeta function.

Now we consider Green's theorem~\eqref{Greenthm} with
\begin{equation}
    A=\frac{1}{\bz-\bar z_2}\ln^2\frac{|z-z_1|^2}{R^2}\ln\frac{|z-z_2|^2}{R^2}\,,\quad B=0\,,
\end{equation}
and find
\begin{equation}
\begin{split}
    & 2\int  \frac{d^2z}{(z-z_1)(\bz-\bar z_2)}\ln\frac{|z-z_1|^2}{R^2}\ln\frac{|z-z_2|^2}{R^2} + \int \frac{d^2z}{(z-z_2)(\bz-\bar z_2)}\ln^2\frac{|z-z_1|^2}{R^2}\\
    &+ \pi\int d^2 z\, \delta^{(2)}(z-z_2)\ln^2\frac{|z-z_1|^2}{R^2}\ln\frac{|z-z_2|^2}{R^2} = 0\,.
 \end{split}
 \end{equation}
Upon using \eqref{logsquare} we find 
\begin{equation}
\label{twologs}
\boxed{
    \int  \frac{d^2z}{(z-z_1)(\bz-\bar z_2)}\ln\frac{|z-z_1|^2}{R^2}\ln\frac{|z-z_2|^2}{R^2} = -\frac{\pi}{3}\ln^3\frac{|z_{12}|^2}{R^2}-2\pi\zeta(3)\,.}
\end{equation}

Finally, we will study the following integral:
\begin{equation}
    \int  \frac{d^2z}{(z-z_1)(\bz-\bar z_2)} I(z_1,z_2,z)\,.
\end{equation}
Consider the Green's theorem~\eqref{Greenthm} with
\begin{equation}
     A=\frac{1}{\bz-\bar z_2}\ln\frac{|z-z_1|^2}{R^2}I(z_1,z_2,z)\,,\quad B=0\,,
\end{equation}
we find
\begin{equation}
\begin{split}
    & \int  \frac{d^2z}{(z-z_1)(\bz-\bar z_2)} I(z_1,z_2,z) + \int \frac{d^2z}{\bz-\bar z_2}\ln\frac{|z-z_1|^2}{R^2}\partial_z I(z_1,z_2,z)\\
    &+ \pi\int d^2z\, \delta^{(2)}(z-z_2)\ln\frac{|z-z_1|^2}{R^2}I(z_1,z_2,z) = 0\,.
 \end{split}
 \end{equation}
A straightforward calculation shows that
\begin{equation}
\boxed{
\begin{split}
     &\int  \frac{d^2z}{(z-z_1)(\bz-\bar z_2)} I(z_1,z_2,z)\\
     &= -\pi\int  \frac{d^2z}{(z-z_1)(\bz-\bar z_2)}\ln\frac{|z-z_1|^2}{R^2}\ln\frac{|z-z_2|^2}{R^2}\\
     &= \frac{\pi^2}{3}\ln^3\frac{|z_{12}|^2}{R^2}+2\pi^2\zeta(3)\,,
     \end{split}}
\end{equation}
where in the last equality we used \eqref{twologs}.
One can then do the $z_1$ and $\bar z_2$ derivative and after some algebra we find
\begin{equation}
\boxed{
    \int \frac{d^2z}{(z-z_1)^2(\bz-\bar z_2)^2} I(z_1,z_2,z) = \frac{2\pi^2}{|z_{12}|^2}  + 
    \pi^3\delta^{(2)}(z_{12})\left(1-\frac12\ln^2\frac{|z_{12}|^2}{R^2}\right)\,.}
\end{equation}

\section{Contractions of structure constants}
\label{appendixB}

In this section, we compute the contractions of structure constants required to evaluate loop corrections to correlation functions. To proceed, we define the matrices $(F_a)_{bc} := f_{abc}$.\footnote{These matrices correspond to the generators $-F_a$ of the algebra in the adjoint representation.} Additionally, we recall that the structure constants are imaginary, i.e., $f_{acd} f_{bcd} = -c_G \delta_{ab}$, and summarize the key properties of the matrices $F_a$ below:
\begin{eqnarray}
    F_aF_a &=&c_G\mathbbm{1}\,,\\
    \Tr(F_aF_b) &=& c_G\delta_{ab}\,,\\
    {}[F_a,F_b] &=& -f_{abc}F_c\,.
\end{eqnarray}
From the above we can easily prove
\begin{equation}
\label{Identity.Fs.adj}
[F_b,[F_b,F_a]]=c_G F_a\,,\quad F_b F_a F_b=\frac12c_GF_a\,.
\end{equation}
Various contractions of structure constants appearing in the computations of the loop-corrections of correlation functions can be written as the traces of products of these matrices. In the following we will list the ones needed in our computations.
\begin{enumerate}
    \item 
    \begin{equation}
        \Tr(F_{a_1}F_{a_3}F_{a_2}F_{a_3}) = \frac{c_G^2}{2}\delta_{a_1a_2}\,.
    \end{equation}
A way to prove this is
    \begin{equation}
    \begin{split}
    \Tr(F_{a_1}F_{a_3}F_{a_2}F_{a_3}) &= \Tr(F_{a_1}F_{a_2}F_{a_3}F_{a_3})+\Tr(F_{a_1}[F_{a_3},F_{a_2}]F_{a_3})\\
    &= c_G\Tr(F_{a_1}F_{a_2})-f_{a_3a_2c}\Tr(F_{a_1}F_cF_{a_3})\\
    &= c_G^2\delta_{a_1a_2}-f_{a_3a_2c}f_{a_1de}f_{cef}f_{a_3fd}\\
    &= c_G^2\delta_{a_1a_2}-f_{a_1ed}f_{fda_3}f_{a_2a_3c}f_{fce}\\
    &= c_G^2\delta_{a_1a_2}-\Tr(F_{a_1}F_{a_3}F_{a_2}F_{a_3})\,.
    \end{split}
    \end{equation}
Alternatively it can be trivially evaluated using \eqref{Identity.Fs.adj}.

    \item 
    \begin{equation}
        \Tr (F_{a_1}F_{a_2}F_{a_3}F_{a_1}F_{a_3}F_{a_2}) = \frac{\dim G c_G^3}{4}\,.
    \end{equation}
A way to prove this is
   \begin{equation}
    \begin{split}
    & \Tr (F_{a_1}F_{a_2}F_{a_3}F_{a_1}F_{a_3}F_{a_2})\\
    &= c_G\Tr (F_{a_1}F_{a_2}F_{a_1}F_{a_2}) + \Tr (F_{a_1}F_{a_2}F_{a_3}[F_{a_1},F_{a_3}]F_{a_2})\\
    &= \frac{\dim G c_G^3}{2} - f_{a_1a_3c}\Tr (F_{a_1}F_{a_2}F_{a_3}F_cF_{a_2})\\
    &= \frac{\dim G c_G^3}{2} - f_{a_1a_3c}f_{a_1de}f_{a_2ef}f_{a_3fg}f_{cgh}f_{a_2hd}\\
    &= \frac{\dim G c_G^3}{2} - f_{da_1e}f_{fea_2}f_{da_2h}f_{chg}f_{fga_3}f_{ca_3a_1}\\
    &= \frac{\dim G c_G^3}{2} - \Tr(F_dF_fF_dF_cF_fF_c)\\
    &= \frac{\dim G c_G^3}{2} - \Tr(F_fF_dF_cF_fF_cF_d)\,.
\end{split}
    \end{equation}
Alternatively it can be trivially evaluated using \eqref{Identity.Fs.adj}.

    \item 
    \begin{equation}
        \Tr(F_{a_1}F_{a_2}F_{a_3}F_{a_1}F_{a_2}F_{a_3}) = 0\,.
    \end{equation}
    The proof to this is
   \begin{equation}
    \begin{split}
         &\Tr(F_{a_1}F_{a_2}F_{a_3}F_{a_1}F_{a_2}F_{a_3})\\
        &= \Tr(F_{a_1}F_{a_2}F_{a_3}F_{a_1}F_{a_3}F_{a_2}) + \Tr(F_{a_1}F_{a_2}F_{a_3}F_{a_1}[F_{a_2},F_{a_3}])\\
        &= \Tr(F_{a_1}F_{a_2}F_{a_3}F_{a_1}F_{a_3}F_{a_2}) - f_{a_2a_3c}\Tr(F_{a_1}F_{a_2}F_{a_3}F_{a_1}F_{c}) \\
        &= \Tr(F_{a_1}F_{a_2}F_{a_3}F_{a_1}F_{a_3}F_{a_2}) - f_{a_2a_3c}f_{a_1de}f_{a_2ef}f_{a_3fg}f_{a_1gh}f_{chd}\\
        &= \Tr(F_{a_1}F_{a_2}F_{a_3}F_{a_1}F_{a_3}F_{a_2}) - f_{eda_1}f_{ha_1g}f_{a_3gf}f_{efa_2}f_{a_3a_2c}f_{hcd}\\
        &= \Tr(F_{a_1}F_{a_2}F_{a_3}F_{a_1}F_{a_3}F_{a_2}) - \Tr(F_{e}F_{h}F_{a_3}F_{e}F_{a_3}F_{h})\\
        &= 0\,.
   \end{split}
    \end{equation}
\end{enumerate}

\section{OO 3-loop}
\label{appendixC}

In this section we will compute in detail the integrals appearing in section \ref{OO3loopint}. There are in total 36 integrals and we can classify them into two types:
\begin{enumerate}
    \item Integral of product of a term from the 5-point function of $J$ and its conjugate in the 5-point function of $\bar J$. There are 6 such integrals.
    \item The other 30 integrals can be paired as one is the complex conjugate of another by picking the complex conjugate of its each factor. 
\end{enumerate}
We will call the first type of integrals the ``diagonal" terms and the second type ``off-diagonal" terms.
\subsection{The ``Diagonal" Terms}
The 6 integrals in the first type can be further classified into 3 pairs where in each pair the two integrals are related to each other by interchanging $z_1$ and $z_2$. So we only need to compute a representative for each pair. The representative of the first pair is
   \begin{equation}
    \begin{split}
    &\frac{(-\lambda)^3}{3!\pi^3k^3}\int d^2 z_{3,4,5} \frac{f_{a_1a_4c}f_{a_2ce}f_{a_3a_5e}f_{a_1a_4f}f_{a_2fg}f_{a_3a_5g}}{z_{12}z_{14}z_{23}z_{35}z_{45}\bz_{12}\bz_{14}\bz_{23}\bz_{35}\bz_{45}}\\
    &= \frac{c_G^3\lambda^3\dim G}{3!\pi^3k^3 |z_{12}|^2}\int \frac{d^2 z_{3,4,5}}{z_{14}z_{23}z_{35}z_{45}\bz_{14}\bz_{23}\bz_{35}\bz_{45}}\,.
   \end{split}
    \end{equation}
The $z_{4,5}$ integral has been computed in section \ref{Section.2loop.composite} where we computed the 2-loop correction to the OO correlator. 
From \eqref{composite.twoloop1} and \eqref{composite.twoloop2} we obtain
\begin{equation}
\int  \frac{d^2 z_{3,4}}{z_{13}z_{24}z_{34}\bz_{13}\bz_{24}\bz_{34}}=\frac{3\pi^2}{|z_{12}|^2}\ln^2\frac{\epsilon^2}{|z_{12}|^2}
\end{equation}
and upon renaming $(z_2,z_3,z_4)\to(z_3,z_4,z_5)$ we find
\begin{equation}
    \int \frac{d^2 z_{4,5} }{z_{14}z_{35}z_{45}\bz_{14}\bz_{35}\bz_{45}} = \frac{3\pi^2}{|z_{13}|^2}\ln^2\frac{\epsilon^2}{|z_{13}|^2}\,.
\end{equation}
Therefore we only need to do the $z_3$ integral
 \begin{equation}
    \begin{split}
    &3\pi^2\int  \frac{d^2 z_3}{z_{13}z_{23}\bz_{13}\bz_{23}}\ln^2\frac{\epsilon^2}{|z_{13}|^2}\\
    &= \frac{3\pi^2}{|z_{12}|^2}\int d^2 z_3 \left(\frac{1}{z_{13}\bz_{13}}-\frac{1}{z_{13}\bz_{23}}-\frac{1}{z_{23}\bz_{13}}+\frac{1}{z_{23}\bz_{23}}\right)\\
    &\qquad\qquad\times\left(\ln^2\frac{\epsilon^2}{R^2} - 2\ln\frac{\epsilon^2}{R^2}\ln\frac{|z_{13}|^2}{R^2} + \ln^2\frac{|z_{13}|^2}{R^2}\right)\,.
   \end{split}
    \end{equation}
One can then use the relevant integrals in appendix \ref{appendix:A} and find it equal to
\begin{equation}
    -\frac{4\pi^3}{|z_{12}|^2}\ln^3\frac{\epsilon^2}{|z_{12}|^2}+\frac{12\pi^3\zeta(3)}{|z_{12}|^2}\,.
\end{equation}
Taking into account the overall factor in front we find
\begin{equation}
    -\frac{2}{3} \frac{c_G^3\lambda^3\dim G}{k^3|z_{12}|^4}\ln^3\frac{\epsilon^2}{|z_{12}|^2}+\frac{2\zeta(3)c_G^3\lambda^3\dim G}{k^3|z_{12}|^4}\,.
\end{equation}
The representative of the second pair is
 \begin{equation}
    \begin{split}
    &\frac{(-\lambda)^3}{3!\pi^3k^3}\int d^2 z_{3,4,5} \frac{f_{a_1a_5c}f_{a_2a_4e}f_{a_3ce}f_{a_1a_5f}f_{a_2a_4g}f_{a_3fg}}{z_{13}z_{15}z_{23}z_{24}z_{45}\bz_{13}\bz_{15}\bz_{23}\bz_{24}\bz_{45}}\\
    &= \frac{c_G^3\lambda^3\dim G}{6\pi^3k^3}\int \frac{d^2 z_{3,4,5} }{z_{13}z_{15}z_{23}z_{24}z_{45}\bz_{13}\bz_{15}\bz_{23}\bz_{24}\bz_{45}}\,.
  \end{split}
    \end{equation}
The integral can be factorized into the product of a $z_{4,5}$ integral
 \begin{equation}
    \begin{split}
    &\int  \frac{d^2 z_{4,5}}{z_{15}z_{24}z_{45}\bz_{15}\bz_{24}\bz_{45}}\\
    &= -2\pi\int \frac{d^2 z_4}{z_{14}z_{24}\bz_{14}\bz_{24}}\ln\frac{\epsilon^2}{|z_{14}|^2}\\
    &= \frac{3\pi^2}{|z_{12}|^2}\ln^2\frac{\epsilon^2}{|z_{12}|^2}
  \end{split}
    \end{equation}
and a $z_3$ integral
\begin{equation}
    \int  \frac{d^2 z_3}{z_{13}z_{23}\bz_{13}\bz_{23}} = -\frac{2\pi}{|z_{12}|^2}\ln\frac{\epsilon^2}{|z_{12}|^2}\,.
\end{equation}
So it is equal to
\begin{equation}
    -\frac{c_G^3\lambda^3\dim G}{k^3|z_{12}|^4}\ln^3\frac{\epsilon^2}{|z_{12}|^2}\,.
\end{equation}
The representative of the third pair is
 \begin{equation}
    \begin{split}
    &\frac{(-\lambda)^3}{3!\pi^3k^3}\int d^2 z_{3,4,5} \frac{f_{a_1a_5c}f_{a_2ce}f_{a_3a_4e}f_{a_1a_5f}f_{a_2fg}f_{a_3a_4g}}{z_{12}z_{15}z_{23}z_{34}z_{45}\bz_{12}\bz_{15}\bz_{23}\bz_{34}\bz_{45}}\\
    &= \frac{c_G^3\lambda^3\dim G}{6\pi^3k^3|z_{12}|^2}\int  \frac{d^2 z_{3,4,5}}{z_{15}z_{23}z_{34}z_{45}\bz_{15}\bz_{23}\bz_{34}\bz_{45}}\\
    &= -\frac{c_G^3\lambda^3\dim G}{3\pi^2k^3|z_{12}|^2}\int \frac{d^2 z_{3,4}}{z_{14}z_{23}z_{34}\bz_{14}\bz_{23}\bz_{34}}\ln\frac{\epsilon^2}{|z_{14}|^2}\\
    &= \frac{c_G^3\lambda^3\dim G}{2\pi k^3|z_{12}|^2}\int  \frac{d^2 z_3}{z_{13}z_{23}\bz_{13}\bz_{23}}\ln^2\frac{\epsilon^2}{|z_{13}|^2}\\
    &= -\frac{2}{3} \frac{c_G^3\lambda^3\dim G}{k^3|z_{12}|^4}\ln^3\frac{\epsilon^2}{|z_{12}|^2}
    +\frac{2\zeta(3)c_G^3\lambda^3\dim G}{k^3|z_{12}|^4}\,.
  \end{split}
    \end{equation}
So the total contribution to the 3-loop correction from the first type of integrals is
\begin{equation}
\label{eq:diagonalcontribution3loop}
\begin{split}
    &2\times\left(-\frac{2}{3}-1-\frac{2}{3}\right)\frac{c_G^3\lambda^3\dim G}{k^3|z_{12}|^4}\ln^3\frac{\epsilon^2}{|z_{12}|^2} 
+2\times(2+2)\frac{\zeta(3)c_G^3\lambda^3\dim G}{k^3|z_{12}|^4}\\
   & = -\frac{14}{3}\frac{c_G^3\lambda^3\dim G}{k^3|z_{12}|^4}\ln^3\frac{\epsilon^2}{|z_{12}|^2}+
   \frac{8\zeta(3)c_G^3\lambda^3\dim G}{k^3|z_{12}|^4}\,.
\end{split}
\end{equation}

\subsection{The ``Off-Diagonal" Terms}
We now turn to the second type of integrals. One can see that the 30 integrals can be classified into 15 pairs and within each pair one is the {\it complex conjugate} of the other. 
Therefore we only need to pick a representative from each pair and compute its {\it real part}. 

One can first use the contraction of the structure constants to show that the contribution from 3 pairs is vanishing. For the remaining 12 pairs as we will show later that the contribution from each pair will be invariant under the interchanging of the external points $z_1$ and $z_2$. Therefore one can further combine integrals related by interchanging $z_1$ and $z_2$ and find that the contribution from the second type of integrals is given in terms of seven terms
\begin{eqnarray}
\label{off.diagonal.OO}
    &&-\frac{\lambda^3c_G^3\dim G}{3!\pi^3k^3}\Re\int d^2 z_{3,4,5}\left(\frac{1}{z_{12}z_{14}z_{23}z_{35}z_{45}\bz_{13}\bz_{15}\bz_{23}\bz_{24}\bz_{45}}+\frac{1}{z_{12}z_{14}z_{23}z_{35}z_{45}\bz_{12}\bz_{13}\bz_{24}\bz_{35}\bz_{45}}\right.\nonumber\\
    &&\qquad\qquad\qquad\qquad\qquad\quad + \frac{2}{z_{12}z_{14}z_{23}z_{35}z_{45}\bz_{13}\bz_{14}\bz_{23}\bz_{25}\bz_{45}} + \frac{1}{z_{12}z_{14}z_{23}z_{35}z_{45}\bz_{12}\bz_{13}\bz_{25}\bz_{34}\bz_{45}}\nonumber\\
    &&\qquad\qquad\qquad\qquad\qquad\quad+\frac{1}{z_{12}z_{15}z_{23}z_{34}z_{45}\bz_{13}\bz_{14}\bz_{23}\bz_{25}\bz_{45}}+\frac{1}{z_{12}z_{15}z_{23}z_{34}z_{45}\bz_{12}\bz_{13}\bz_{25}\bz_{34}\bz_{45}}\nonumber\\
    &&\left.\qquad\qquad\qquad\qquad\qquad\quad + \frac{2}{z_{12}z_{15}z_{23}z_{34}z_{45}\bz_{13}\bz_{15}\bz_{23}\bz_{24}\bz_{45}}\right)\,.
\end{eqnarray}
We will now compute them one by one.
\begin{enumerate}
    \item \begin{equation}
        \Re\int  \frac{d^2 z_{3,4,5}}{z_{12}z_{14}z_{23}z_{35}z_{45}\bz_{13}\bz_{15}\bz_{23}\bz_{24}\bz_{45}}\,.
    \end{equation}
    We first integrate over $z_5$ and obtain
    \begin{equation}
        \int  \frac{d^2 z_5}{z_{35}z_{45}\bz_{15}\bz_{45}} = -\frac{\pi}{z_{34}\bz_{14}}\left(\ln\frac{\epsilon^2}{R^2}+\ln\frac{|z_{13}|^2}{R^2}-\ln\frac{|z_{14}|^2}{R^2}-\ln\frac{|z_{34}|^2}{R^2}\right)\,.
    \end{equation}
    So the integral becomes
    \begin{eqnarray}
    \label{first.off.diagonal}
        &&-\pi\Re\int  \frac{d^2 z_{3,4}}{z_{12}z_{14}z_{23}z_{34}\bz_{13}\bz_{14}\bz_{23}\bz_{24}}\left(\ln\frac{\epsilon^2}{R^2}+\ln\frac{|z_{13}|^2}{R^2}-\ln\frac{|z_{14}|^2}{R^2}-\ln\frac{|z_{34}|^2}{R^2}\right)\nonumber\\
        &=& -\pi\int \frac{d^2 z_3}{ |z_{12}|^2} \left[\frac{1}{z_{13}z_{23}\bz_{13}\bz_{23}}\Re\int d^2 z_4 \left(\frac{1}{z_{14}\bz_{14}}-\frac{1}{z_{14}\bz_{24}}-\frac{1}{z_{34}\bz_{14}}+\frac{1}{z_{34}\bz_{24}}\right)\right.\nonumber\\
        &&\qquad\qquad\qquad\qquad\qquad\left.\times\left(\ln\frac{\epsilon^2}{R^2}+\ln\frac{|z_{13}|^2}{R^2}-\ln\frac{|z_{14}|^2}{R^2}-\ln\frac{|z_{34}|^2}{R^2}\right)\right]\,.
    \end{eqnarray}
    The relevant integrals in the $z_4$ integral have been computed in appendix \ref{appendix:A} and thus we will find that this gives
    \begin{equation}
        \frac{\pi^2}{2|z_{12}|^2}\int  \frac{d^2 z_3}{z_{13}z_{23}\bz_{13}\bz_{23}} \left(\ln^2\frac{\epsilon^2}{|z_{12}|^2}+\ln^2\frac{\epsilon^2}{|z_{13}|^2}-\ln^2\frac{\epsilon^2}{|z_{23}|^2}\right)\,.
    \end{equation}
    As one can see in the computation of the first representative of the ``diagonal" terms the second and third terms in the equation above will cancel each other. Therefore we will get in the end
    \begin{equation}
    \label{first.off.diagonal.result}
        \Re\int \frac{d^2 z_{3,4,5}}{z_{12}z_{14}z_{23}z_{35}z_{45}\bz_{13}\bz_{15}\bz_{23}\bz_{24}\bz_{45}} = -\frac{\pi^3}{|z_{12}|^4}\ln^3\frac{\epsilon^2}{|z_{12}|^2}\,.
    \end{equation}

    \item \begin{equation}
        \Re\int  \frac{d^2 z_{3,4,5}}{z_{12}z_{14}z_{23}z_{35}z_{45}\bz_{12}\bz_{13}\bz_{24}\bz_{35}\bz_{45}}\,.
    \end{equation}
    We integrate over $z_5$ first and find this equal to
    \begin{eqnarray}
        &&-\frac{2\pi}{|z_{12}|^2}\Re\int  \frac{d^2 z_{3,4}}{z_{14}z_{23}z_{34}\bz_{13}\bz_{24}\bz_{34}}\ln\frac{\epsilon^2}{|z_{34}|^2}\\
        &=& -\frac{2\pi}{|z_{12}|^2}\int \frac{d^2 z_3}{z_{13}z_{23}\bz_{13}\bz_{23}} \Re \left[ \int d^2 z_4 \left(\frac{1}{z_{14}\bz_{24}} - \frac{1}{z_{14}\bz_{34}} - \frac{1}{z_{34}\bz_{24}} + \frac{1}{z_{34}\bz_{34}}\right)\ln\frac{\epsilon^2}{|z_{34}|^2}\right]\,.\nonumber
    \end{eqnarray}
    We now focus on the $z_4$ integral
\begin{equation}
\label{non.diagonal.second0}
\begin{split}
        &\Re \int d^2 z_4 \left(\frac{1}{z_{14}\bz_{24}} - \frac{1}{z_{14}\bz_{34}} - \frac{1}{z_{34}\bz_{24}} + \frac{1}{z_{34}\bz_{34}}\right)\ln\frac{\epsilon^2}{|z_{34}|^2}\\
        &= \Re \int d^2 z_4 \left(\frac{1}{z_{14}\bz_{24}} - \frac{1}{z_{14}\bz_{34}} - \frac{1}{z_{34}\bz_{24}} + \frac{1}{z_{34}\bz_{34}}\right)\left(\ln\frac{\epsilon^2}{R^2}-\ln\frac{|z_{34}|^2}{R^2}\right)\\
        &= -\pi\ln\frac{\epsilon^2}{R^2}\left(\ln\frac{\epsilon^2}{R^2}+\ln\frac{|z_{12}|^2}{R^2}-\ln\frac{|z_{13}|^2}{R^2}-\ln\frac{|z_{23}|^2}{R^2}\right)\\
        &+\frac{\pi}{2}\left(\ln\frac{|z_{12}|^2}{R^2}\ln\frac{|z_{13}|^2}{R^2}+\ln\frac{|z_{12}|^2}{R^2}\ln\frac{|z_{23}|^2}{R^2}-\ln\frac{|z_{13}|^2}{R^2}\ln\frac{|z_{23}|^2}{R^2}\right)\\
        &-\frac{\pi}{2}\left(\ln^2\frac{|z_{13}|^2}{R^2}+\ln^2\frac{|z_{23}|^2}{R^2}-\ln^2\frac{\epsilon^2}{R^2}\right)\\
        &= \frac{\pi}{2}\left(-\ln^2\frac{\epsilon^2}{|z_{13}|^2}-\ln^2\frac{\epsilon^2}{|z_{23}|^2}+\ln^2\frac{\epsilon^2}{|z_{12}|^2}-\ln^2\frac{|z_{12}|^2}{R^2}\right.\\
        &+\left.\ln\frac{|z_{12}|^2}{R^2}\ln\frac{|z_{13}|^2}{R^2}+\ln\frac{|z_{12}|^2}{R^2}\ln\frac{|z_{23}|^2}{R^2}-\ln\frac{|z_{13}|^2}{R^2}\ln\frac{|z_{23}|^2}{R^2}\right)\,.
     \end{split}
    \end{equation}
    Now we need to compute the $z_3$ integral
 \begin{equation}
\begin{split}
\label{non.diagonal.second}
        & -\frac{\pi^2}{|z_{12}|^2}\int  \frac{d^2 z_3}{z_{13}z_{23}\bz_{13}\bz_{23}}\left(-\ln^2\frac{\epsilon^2}{|z_{13}|^2}-\ln^2\frac{\epsilon^2}{|z_{23}|^2}+\ln^2\frac{\epsilon^2}{|z_{12}|^2}-\ln^2\frac{|z_{12}|^2}{R^2}\right.\\
        &\qquad\qquad +\left.\ln\frac{|z_{12}|^2}{R^2}\ln\frac{|z_{13}|^2}{R^2}+\ln\frac{|z_{12}|^2}{R^2}\ln\frac{|z_{23}|^2}{R^2}-\ln\frac{|z_{13}|^2}{R^2}\ln\frac{|z_{23}|^2}{R^2}\right)\,.
   \end{split}
    \end{equation}
    All the relevant integrals have been previously computed and thus we find
    \begin{equation}
    \label{non.diagonal.second.f}
        \Re\int \frac{d^2 z_{3,4,5}}{z_{12}z_{14}z_{23}z_{35}z_{45}\bz_{12}\bz_{13}\bz_{24}\bz_{35}\bz_{45}} = -\frac{2\pi^3}{3|z_{12}|^4}\ln^3\frac{\epsilon^2}{|z_{12}|^2}
        +\frac{12\zeta(3)\pi^3}{|z_{12}|^4}\,.
    \end{equation}

    \item \begin{equation}
        \Re \int  \frac{d^2 z_{3,4,5}}{z_{12}z_{14}z_{23}z_{35}z_{45}\bz_{13}\bz_{14}\bz_{23}\bz_{25}\bz_{45}}\,.
    \end{equation}
    We integrate over $z_5$ first and find
    \begin{equation}
        \int \frac{d^2 z_5 }{z_{35}z_{45}\bz_{25}\bz_{45}} = -\frac{\pi}{z_{34}\bz_{24}}\left(\ln\frac{\epsilon^2}{R^2}+\ln\frac{|z_{23}|^2}{R^2}-\ln\frac{|z_{24}|^2}{R^2}-\ln\frac{|z_{34}|^2}{R^2}\right)\,.
    \end{equation}
    Then we need to compute
 \begin{equation}
\begin{split}
        &-\frac{\pi}{|z_{12}|^2}\int \frac{ d^2 z_3}{z_{13}z_{23}\bz_{13}\bz_{23}}\Re \left[\int d^2 z_4 \left(\frac{1}{z_{14}\bz_{14}}-\frac{1}{z_{14}\bz_{24}}-\frac{1}{z_{34}\bz_{14}}+\frac{1}{z_{34}\bz_{24}}\right) \right.\\
        &\qquad\qquad\qquad\qquad\qquad\left.\times\left(\ln\frac{\epsilon^2}{R^2}+\ln\frac{|z_{23}|^2}{R^2}-\ln\frac{|z_{24}|^2}{R^2}-\ln\frac{|z_{34}|^2}{R^2}\right)\right]\,,
   \end{split}
    \end{equation}    
where we find    
 \begin{equation}
 \label{non.diagonal.third}
\begin{split}
        &-\Re \left[\int d^2 z_4 \left(\frac{1}{z_{14}\bz_{14}}-\frac{1}{z_{14}\bz_{24}}-\frac{1}{z_{34}\bz_{14}}+\frac{1}{z_{34}\bz_{24}}\right) \right.\\
        &\qquad\qquad\qquad\qquad\qquad\left.\times\left(\ln\frac{\epsilon^2}{R^2}+\ln\frac{|z_{23}|^2}{R^2}-\ln\frac{|z_{24}|^2}{R^2}-\ln\frac{|z_{34}|^2}{R^2}\right)\right]\\
    &=\pi\left(\ln^2\frac{\epsilon^2}{|z_{12}|^2} - 2\ln\frac{\epsilon^2}{R^2}\ln\frac{|z_{13}|^2}{|z_{23}|^2}+\ln\frac{|z_{12}|^2}{R^2}\ln\frac{|z_{13}|^2}{|z_{23}|^2}\right.\\
        &\left.\qquad\qquad\qquad-\ln\frac{|z_{13}|^2}{R^2}\ln\frac{|z_{23}|^2}{R^2}+\ln^2\frac{|z_{13}|^2}{R^2}\right)\,.
   \end{split}
    \end{equation}
This will give
   \begin{equation}
   \label{non.diagonal.third1}
\begin{split}
        &\frac{\pi^2}{|z_{12}|^2}\int  \frac{d^2 z_3}{z_{13}z_{23}\bz_{13}\bz_{23}}\left(\ln^2\frac{\epsilon^2}{|z_{12}|^2} - 2\ln\frac{\epsilon^2}{R^2}\ln\frac{|z_{13}|^2}{|z_{23}|^2}+\ln\frac{|z_{12}|^2}{R^2}\ln\frac{|z_{13}|^2}{|z_{23}|^2}\right.\\
        &\qquad\qquad\qquad\left.-\ln\frac{|z_{13}|^2}{R^2}\ln\frac{|z_{23}|^2}{R^2}+\ln^2\frac{|z_{13}|^2}{R^2}\right)\,.
    \end{split}
    \end{equation}
    One can use the relevant integrals computed previously to find that the second and third terms in the above expression both give zero contribution while the remaining terms give us
    \begin{equation}
    \label{non.diagonal.third.f}
        \Re \int  \frac{d^2 z_{3,4,5}}{z_{12}z_{14}z_{23}z_{35}z_{45}\bz_{13}\bz_{14}\bz_{23}\bz_{25}\bz_{45}} = -\frac{7\pi^3}{3|z_{12}|^4}\ln^3\frac{\epsilon^2}{|z_{12}|^2}\,.
    \end{equation}

    \item
    \begin{equation}
        \Re\int \frac{d^2 z_{3,4,5} }{z_{12}z_{14}z_{23}z_{35}z_{45}\bz_{12}\bz_{13}\bz_{25}\bz_{34}\bz_{45}}\,.
    \end{equation}
    We integrate over $z_5$ first and find
    \begin{equation}
        \int  \frac{d^2 z_5}{z_{35}z_{45}\bz_{25}\bz_{45}} = -\frac{\pi}{z_{34}\bz_{24}}\left(\ln\frac{\epsilon^2}{R^2}+\ln\frac{|z_{23}|^2}{R^2}-\ln\frac{|z_{24}|^2}{R^2}-\ln\frac{|z_{34}|^2}{R^2}\right)\,.
    \end{equation}
    So we now have
    \begin{equation}
\begin{split}
        & -\frac{\pi}{|z_{12}|^2}\int  \frac{d^2 z_3}{z_{13}z_{23}\bz_{13}\bz_{23}} \Re\left[\int d^2 z_4 \left(\frac{1}{z_{14}\bz_{24}} - \frac{1}{z_{14}\bz_{34}} - \frac{1}{z_{34}\bz_{24}} + \frac{1}{z_{34}\bz_{34}}\right)\right.\\
        &\qquad\qquad\qquad\qquad\qquad\left.\times\left(\ln\frac{\epsilon^2}{R^2}+\ln\frac{|z_{23}|^2}{R^2}-\ln\frac{|z_{24}|^2}{R^2}-\ln\frac{|z_{34}|^2}{R^2}\right)\right]\,.
      \end{split}
    \end{equation}
    We focus on the $z_4$ integral.
     \begin{equation}
\begin{split}
        &\Re\int d^2 z_4 \left(\frac{1}{z_{14}\bz_{24}} - \frac{1}{z_{14}\bz_{34}} - \frac{1}{z_{34}\bz_{24}} + \frac{1}{z_{34}\bz_{34}}\right)\\
        &\qquad\quad\times\left(\ln\frac{\epsilon^2}{R^2}+\ln\frac{|z_{23}|^2}{R^2}-\ln\frac{|z_{24}|^2}{R^2}-\ln\frac{|z_{34}|^2}{R^2}\right)\\
        &= -\pi\left(\ln\frac{\epsilon^2}{R^2}+\ln\frac{|z_{23}|^2}{R^2}\right)\left(\ln\frac{\epsilon^2}{R^2}+\ln\frac{|z_{12}|^2}{R^2}-\ln\frac{|z_{13}|^2}{R^2}-\ln\frac{|z_{23}|^2}{R^2}\right)\\
        &-\Re\int d^2 z_4\left(\frac{1}{z_{14}\bz_{24}} - \frac{1}{z_{14}\bz_{34}} - \frac{1}{z_{34}\bz_{24}} + \frac{1}{z_{34}\bz_{34}}\right)\left(\ln\frac{|z_{24}|^2}{R^2}+\ln\frac{|z_{34}|^2}{R^2}\right)\\
        &= -\pi\left(\ln\frac{\epsilon^2}{R^2}+\ln\frac{|z_{23}|^2}{R^2}\right)\left(\ln\frac{\epsilon^2}{R^2}+\ln\frac{|z_{12}|^2}{R^2}-\ln\frac{|z_{13}|^2}{R^2}-\ln\frac{|z_{23}|^2}{R^2}\right)\\
        & +\frac{\pi}{2}\ln^2\frac{|z_{12}|^2}{R^2}-\frac{\pi}{2}\ln\frac{|z_{12}|^2}{R^2}\ln\frac{|z_{13}|^2}{R^2}+\frac{\pi}{2}\ln\frac{|z_{12}|^2}{R^2}\ln\frac{|z_{23}|^2}{R^2}\\
        &-\frac{\pi}{2}\ln\frac{|z_{13}|^2}{R^2}\ln\frac{|z_{23}|^2}{R^2}-\frac{\pi}{2}\ln^2\frac{|z_{23}|^2}{R^2}+\pi\ln\frac{\epsilon^2}{R^2}\ln\frac{|z_{23}|^2}{R^2}-\frac{\pi}{2}\ln^2\frac{|z_{23}|^2}{R^2}\\
        &+\frac{\pi}{2}\ln\frac{|z_{12}|^2}{R^2}\ln\frac{|z_{13}|^2}{R^2}+\frac{\pi}{2}\ln\frac{|z_{12}|^2}{R^2}\ln\frac{|z_{23}|^2}{R^2}-\frac{\pi}{2}\ln\frac{|z_{13}|^2}{R^2}\ln\frac{|z_{23}|^2}{R^2}\\
        & -\frac{\pi}{2}\ln^2\frac{|z_{13}|^2}{R^2}-\frac{\pi}{2}\ln^2\frac{|z_{23}|^2}{R^2}+\frac{\pi}{2}\ln^2\frac{\epsilon^2}{R^2}\\
        &= \frac{\pi}{2}\left(\ln^2\frac{\epsilon^2}{|z_{12}|^2}-\ln^2\frac{\epsilon^2}{|z_{13}|^2}-\ln^2\frac{\epsilon^2}{|z_{23}|^2}\right)\,.
    \end{split}
    \end{equation}
    Finally we do the $z_3$ integral and find
     \begin{equation}
\begin{split}
        &\Re\int \frac{d^2 z_{3,4,5} }{z_{12}z_{14}z_{23}z_{35}z_{45}\bz_{12}\bz_{13}\bz_{25}\bz_{34}\bz_{45}}\\
        &= -\frac{\pi^2}{2|z_{12}|^2}\int  \frac{d^2 z_3}{z_{13}z_{23}\bz_{13}\bz_{23}}\left(\ln^2\frac{\epsilon^2}{|z_{12}|^2}-\ln^2\frac{\epsilon^2}{|z_{13}|^2}-\ln^2\frac{\epsilon^2}{|z_{23}|^2}\right)\\
        &= -\frac{\pi^3}{3|z_{12}|^4}\ln^3\frac{\epsilon^2}{|z_{12}|^2}+\frac{4\zeta(3)\pi^3}{|z_{12}|^4}\,.
  \end{split}
    \end{equation}

    \item \begin{equation}
        \Re\int  \frac{d^2 z_{3,4,5}}{z_{12}z_{15}z_{23}z_{34}z_{45}\bz_{13}\bz_{14}\bz_{23}\bz_{25}\bz_{45}}\,.
    \end{equation}
    We integrate over $z_5$ first and find
    \begin{equation}
        \int  \frac{d^2 z_5}{z_{15}z_{45}\bz_{25}\bz_{45}} = -\frac{\pi}{z_{14}\bz_{24}}\left(\ln\frac{\epsilon^2}{R^2}+\ln\frac{|z_{12}|^2}{R^2}-\ln\frac{|z_{14}|^2}{R^2}-\ln\frac{|z_{24}|^2}{R^2}\right)\,.
    \end{equation}
    So we need to compute
       \begin{equation}
\begin{split}
        &-\pi\Re\int  \frac{d^2 z_{3,4}}{z_{12}z_{14}z_{23}z_{34}\bz_{13}\bz_{14}\bz_{23}\bz_{24}}\left(\ln\frac{\epsilon^2}{R^2}+\ln\frac{|z_{12}|^2}{R^2}-\ln\frac{|z_{14}|^2}{R^2}-\ln\frac{|z_{24}|^2}{R^2}\right)\\
        &= -\frac{\pi}{|z_{12}|^2}\int \frac{d^2 z_3}{z_{13}z_{23}\bz_{13}\bz_{23}}\Re \left[\int d^2 z_4 \left(\frac{1}{z_{14}\bz_{14}}-\frac{1}{z_{14}\bz_{24}}-\frac{1}{z_{34}\bz_{14}}+\frac{1}{z_{34}\bz_{24}}\right)\right.\\
        &\qquad\qquad\qquad\qquad\qquad\left.\times\left(\ln\frac{\epsilon^2}{R^2}+\ln\frac{|z_{12}|^2}{R^2}-\ln\frac{|z_{14}|^2}{R^2}-\ln\frac{|z_{24}|^2}{R^2}\right)\right]\,.
     \end{split}
    \end{equation}
We can see that the $z_4$ integral can be acquired from \eqref{first.off.diagonal}, obtained in the analysis of the first term in \eqref{off.diagonal.OO}, by interchanging $z_2$ and $z_3$ and then take complex conjugate. Since we are only taking the real part, it suffices to interchanging $z_2$ and $z_3$ in the result there which is already symmetric in $z_2$ and $z_3$. We can then see that this term is equal to the first term in \eqref{off.diagonal.OO}, was given in \eqref{first.off.diagonal.result}
    \begin{equation}
        \Re\int  \frac{d^2 z_{3,4,5}}{z_{12}z_{15}z_{23}z_{34}z_{45}\bz_{13}\bz_{14}\bz_{23}\bz_{25}\bz_{45}} = -\frac{\pi^3}{|z_{12}|^4}\ln^3\frac{\epsilon^2}{|z_{12}|^2}\,.
    \end{equation}

    \item
    \begin{equation}
        \Re\int  \frac{d^2 z_{3,4,5}}{z_{12}z_{15}z_{23}z_{34}z_{45}\bz_{12}\bz_{13}\bz_{25}\bz_{34}\bz_{45}}\,.
    \end{equation}
    We first integrate over $z_5$ and find
    \begin{equation}
        \int  \frac{d^2 z_5}{z_{15}z_{45}\bz_{25}\bz_{45}} = -\frac{\pi}{z_{14}\bz_{24}}\left(\ln\frac{\epsilon^2}{R^2}+\ln\frac{|z_{12}|^2}{R^2}-\ln\frac{|z_{14}|^2}{R^2}-\ln\frac{|z_{24}|^2}{R^2}\right)\,.
    \end{equation}
    So we need to compute
       \begin{equation}
\begin{split}
        &-\frac{\pi}{|z_{12}|^2}\Re\int  \frac{d^2 z_{3,4}}{z_{14}z_{23}z_{34}\bz_{13}\bz_{24}\bz_{34}}\left(\ln\frac{\epsilon^2}{R^2}+\ln\frac{|z_{12}|^2}{R^2}-\ln\frac{|z_{14}|^2}{R^2}-\ln\frac{|z_{24}|^2}{R^2}\right)\\
        &= -\frac{\pi}{|z_{12}|^2}\int  \frac{d^2 z_3}{z_{13}z_{23}\bz_{13}\bz_{23}}\left[\Re\int d^2 z_4 \left(\frac{1}{z_{14}\bz_{24}} - \frac{1}{z_{14}\bz_{34}} - \frac{1}{z_{34}\bz_{24}} + \frac{1}{z_{34}\bz_{34}}\right)\right.\\
        &\qquad\qquad\qquad\qquad\qquad\left.\times\left(\ln\frac{\epsilon^2}{R^2}+\ln\frac{|z_{12}|^2}{R^2}-\ln\frac{|z_{14}|^2}{R^2}-\ln\frac{|z_{24}|^2}{R^2}\right)\right]\,.
  \end{split}
    \end{equation}
    The $z_4$ integral can be acquired from \eqref{non.diagonal.third}, obtained in the analysis of the third term in \eqref{off.diagonal.OO}, by interchanging $z_1$ and $z_3$, so we find
       \begin{equation}
\begin{split}
        &\Re \int d^2 z_4 \left(\frac{1}{z_{34}\bz_{34}}-\frac{1}{z_{34}\bz_{24}}-\frac{1}{z_{14}\bz_{34}}+\frac{1}{z_{14}\bz_{24}}\right) \\
        &\times\left(\ln\frac{\epsilon^2}{R^2}+\ln\frac{|z_{12}|^2}{R^2}-\ln\frac{|z_{24}|^2}{R^2}-\ln\frac{|z_{14}|^2}{R^2}\right)\\
        &= -\pi\left(\ln^2\frac{\epsilon^2}{|z_{23}|^2} - 2\ln\frac{\epsilon^2}{R^2}\ln\frac{|z_{13}|^2}{|z_{12}|^2}+\ln\frac{|z_{23}|^2}{R^2}\ln\frac{|z_{13}|^2}{|z_{12}|^2}\right.\\
        &\qquad\left.-\ln\frac{|z_{13}|^2}{R^2}\ln\frac{|z_{12}|^2}{R^2}+\ln^2\frac{|z_{13}|^2}{R^2}\right)\\
        &= \pi\left(-\ln^2\frac{\epsilon^2}{|z_{13}|^2}-\ln^2\frac{\epsilon^2}{|z_{23}|^2}+\ln^2\frac{\epsilon^2}{|z_{12}|^2}-\ln^2\frac{|z_{12}|^2}{R^2}\right.\\
        &\quad+\left.\ln\frac{|z_{12}|^2}{R^2}\ln\frac{|z_{13}|^2}{R^2}+\ln\frac{|z_{12}|^2}{R^2}\ln\frac{|z_{23}|^2}{R^2}-\ln\frac{|z_{13}|^2}{R^2}\ln\frac{|z_{23}|^2}{R^2}\right)\,.
  \end{split}
    \end{equation}
It is then found that the $z_3$ integral is identical to the one in \eqref{non.diagonal.second}, which corresponds to the second term of \eqref{off.diagonal.OO}, as given by \eqref{non.diagonal.second.f}    
\begin{equation}
        \Re\int  \frac{d^2 z_{3,4,5}}{z_{12}z_{15}z_{23}z_{34}z_{45}\bz_{12}\bz_{13}\bz_{25}\bz_{34}\bz_{45}} = -\frac{2\pi^3}{3|z_{12}|^4}\ln^3\frac{\epsilon^2}{|z_{12}|^2}
       +\frac{12\zeta(3)\pi^3}{|z_{12}|^4}\,.
    \end{equation}

    \item 
    \begin{equation}
        \Re\int \frac{d^2 z_{3,4,5} }{z_{12}z_{15}z_{23}z_{34}z_{45}\bz_{13}\bz_{15}\bz_{23}\bz_{24}\bz_{45}}\,.
    \end{equation}
    We integrate over $z_5$ first and find this equal to
    \begin{eqnarray}
        && -2\pi\int \frac{d^2 z_{3,4}}{z_{12}z_{23}z_{14}z_{34}\bz_{13}\bz_{23}\bz_{14}\bz_{24}}\ln\frac{\epsilon^2}{|z_{14}|^2}\\
        &=& -\frac{2\pi}{|z_{12}|^2}\int  \frac{d^2 z_3}{z_{13}z_{23}\bz_{13}\bz_{23}} \Re\left[ \int d^2 z_4 \left(\frac{1}{z_{14}\bz_{14}} - \frac{1}{z_{14}\bz_{24}} - \frac{1}{z_{34}\bz_{14}} + \frac{1}{z_{34}\bz_{24}}\right)\ln\frac{\epsilon^2}{|z_{14}|^2}\right]\nonumber\,.
    \end{eqnarray}
    The $z_4$ integral can be acquired from \eqref{non.diagonal.second0}, in the second term in \eqref{off.diagonal.OO}, by interchanging $z_1$ and $z_3$ and thus
     \begin{equation}
\begin{split}
        &\Re \int d^2 z_4 \left(\frac{1}{z_{14}\bz_{14}} - \frac{1}{z_{14}\bz_{24}} - \frac{1}{z_{34}\bz_{14}} + \frac{1}{z_{34}\bz_{24}}\right)\ln\frac{\epsilon^2}{|z_{14}|^2}\\
        &= \frac{\pi}{2}\left(-\ln^2\frac{\epsilon^2}{|z_{13}|^2}-\ln^2\frac{\epsilon^2}{|z_{12}|^2}+\ln^2\frac{\epsilon^2}{|z_{23}|^2}-\ln^2\frac{|z_{23}|^2}{R^2}\right.\\
        &+\left.\ln\frac{|z_{23}|^2}{R^2}\ln\frac{|z_{13}|^2}{R^2}+\ln\frac{|z_{23}|^2}{R^2}\ln\frac{|z_{12}|^2}{R^2}-\ln\frac{|z_{13}|^2}{R^2}\ln\frac{|z_{12}|^2}{R^2}\right)\\
        &= -\frac{\pi}{2}\left(\ln^2\frac{\epsilon^2}{|z_{12}|^2} - 2\ln\frac{\epsilon^2}{R^2}\ln\frac{|z_{13}|^2}{|z_{23}|^2}+\ln\frac{|z_{12}|^2}{R^2}\ln\frac{|z_{13}|^2}{|z_{23}|^2}\right.\\
        &\qquad\left.-\ln\frac{|z_{13}|^2}{R^2}\ln\frac{|z_{23}|^2}{R^2}+\ln^2\frac{|z_{13}|^2}{R^2}\right)\,.
  \end{split}
    \end{equation}
    Then one can find that the $z_3$ integral is same as in \eqref{non.diagonal.third1}, which corresponds to the third term in \eqref{off.diagonal.OO}, given by \eqref{non.diagonal.third.f}
    \begin{equation}
        \Re\int  \frac{d^2 z_{3,4,5}}{z_{12}z_{15}z_{23}z_{34}z_{45}\bz_{13}\bz_{15}\bz_{23}\bz_{24}\bz_{45}} = -\frac{7\pi^3}{3|z_{12}|^4}\ln^3\frac{\epsilon^2}{|z_{12}|^2}\,.
    \end{equation}
\end{enumerate}
So the total contribution to the 3-loop correction from the second type of integrals is
\begin{equation}
\begin{split}
&\frac{\lambda^3c_G^3\dim G}{3!k^3|z_{12}|^4}\ln^3\frac{\epsilon^2}{|z_{12}|^2}\left(2+\frac{4}{3}+\frac{28}{3}+\frac{1}{3}\right) 
-\frac{\zeta(3)\lambda^3c_G^3\dim G}{3!k^3|z_{12}|^4}\left(12+4+12\right)\\
&= \frac{13\lambda^3c_G^3\dim G}{6k^3|z_{12}|^4}\ln^3\frac{\epsilon^2}{|z_{12}|^2}-\frac{14\zeta(3)\lambda^3c_G^3\dim G}{3k^3|z_{12}|^4}\,.
\end{split}
\end{equation}
Adding this to the results from the first type of integrals \eqref{eq:diagonalcontribution3loop} we find
\begin{equation}
   \langle O(z_1)O(z_2)\rangle^{(3)}=   -\frac{5\lambda^3c_G^3\dim G}{2k^3|z_{12}|^4}\ln^3\frac{\epsilon^2}{|z_{12}|^2}
   +\frac{10\zeta(3)\lambda^3c_G^3\dim G}{3k^3|z_{12}|^4}\,.
\end{equation}

\section{JJ 3-loop}
\label{appendixD}
In this section we will compute in detail the 3 integrals in equation \eqref{JJ3loopint}.
\begin{itemize}
    \item The first integral is
    \begin{equation}
      \int \frac{d^2 z_{3,4,5} }{z_{12}z_{13}z_{25}z_{34}z_{45}\bz_{34}\bz_{35}\bz_{45}}\,.
    \end{equation}
    First we integrate over $z_5$ and have
    \begin{equation}
        \int  \frac{d^2 z_5}{z_{25}z_{45}\bz_{35}\bz_{45}} = -\frac{\pi}{z_{24}\bz_{34}}\left( \ln\frac{|z_{23}|^2}{R^2} + \ln\frac{\epsilon^2}{R^2} - \ln\frac{|z_{24}|^2}{R^2} - \ln\frac{|z_{34}|^2}{R^2}\right)\,.
    \end{equation}
    Now we need to compute the $z_4$ integral
      \begin{equation}
\begin{split}
        &-\pi\int  \frac{d^2 z_4}{z_{24}z_{34}\bz_{34}^2}\left( \ln\frac{|z_{23}|^2}{R^2} + \ln\frac{\epsilon^2}{R^2} - \ln\frac{|z_{24}|^2}{R^2} - \ln\frac{|z_{34}|^2}{R^2}\right)\\
        &= \frac{\pi}{z_{23}}\int d^2 z_4 \left(\frac{1}{z_{24}\bz_{34}^2}-\frac{1}{z_{34}\bz_{34}^2} \right)\left( \ln\frac{|z_{23}|^2}{R^2} + \ln\frac{\epsilon^2}{R^2} - \ln\frac{|z_{24}|^2}{R^2} - \ln\frac{|z_{34}|^2}{R^2}\right)\\
        &= \frac{\pi}{z_{23}}\int d^2 z_4 \left[\frac{1}{z_{24}\bz_{34}^2} \left( \ln\frac{|z_{23}|^2}{R^2} + \ln\frac{\epsilon^2}{R^2}\right) -\frac{1}{z_{24}\bz_{34}^2}\ln\frac{|z_{24}|^2}{R^2}\right.\\
        &\left.\qquad\qquad\quad - \frac{1}{z_{24}\bz_{34}^2}\ln\frac{|z_{34}|^2}{R^2} + \frac{1}{z_{34}\bz_{34}^2}\ln\frac{|z_{24}|^2}{R^2}\right]\\
        &= \frac{\pi^2}{z_{23}\bz_{23}}\left[-\left( \ln\frac{|z_{23}|^2}{R^2} + \ln\frac{\epsilon^2}{R^2}\right) +\ln\frac{|z_{23}|^2}{R^2} +2 +2\ln\frac{|z_{23}|^2}{R^2} -\ln\frac{\epsilon^2}{R^2} \right]\\
        &= \frac{2\pi^2}{z_{23}\bz_{23}}\left(-\ln\frac{\epsilon^2}{|z_{23}|^2}+1\right)\,.
        \end{split}
\end{equation}
    Finally we do the $z_3$ integral and find the first integral equals to
    \begin{equation}
    \label{JJ.3loop.first}
\begin{split}
        & 2\pi^2\int  \frac{d^2 z_3}{z_{12}z_{13}z_{23}\bz_{23}}\left(-\ln\frac{\epsilon^2}{|z_{23}|^2}+1\right)\\
        &= \frac{2\pi^2}{z_{12}^2} \int d^2 z_3 \left(\frac{1}{z_{23}\bz_{23}}-\frac{1}{z_{13}\bz_{23}}\right)\left(1-\ln\frac{\epsilon^2}{R^2}+\ln\frac{|z_{23}|^2}{R^2}\right)\\
        &= -\frac{\pi^3}{z_{12}^2}\left[2\ln\frac{\epsilon^2}{|z_{12}|^2}\left(1-\ln\frac{\epsilon^2}{R^2}\right)+\ln^2\frac{\epsilon^2}{R^2}-\ln^2\frac{|z_{12}|^2}{R^2}\right]\\
        &= \frac{\pi^3}{z_{12}^2}\left(\ln^2\frac{\epsilon^2}{|z_{12}|^2}-2\ln\frac{\epsilon^2}{|z_{12}|^2}\right)\,,
       \end{split}
\end{equation}
So we find
\begin{equation}
   \int \frac{d^2 z_{3,4,5} }{z_{12}z_{13}z_{25}z_{34}z_{45}\bz_{34}\bz_{35}\bz_{45}}=\frac{\pi^3}{z_{12}^2}\left(\ln^2\frac{\epsilon^2}{|z_{12}|^2}-2\ln\frac{\epsilon^2}{|z_{12}|^2}\right)\,.
\end{equation}

    \item The second integral is
    \begin{equation}
        \int  \frac{d^2 z_{3,4,5}}{z_{12}z_{13}z_{24}z_{35}z_{45}\bz_{34}\bz_{35}\bz_{45}}\,.
    \end{equation}
    First we integrate over $z_5$ and have
    \begin{equation}
        \int  \frac{d^2 z_5}{z_{35}z_{45}\bz_{35}\bz_{45}} = -\frac{2\pi}{|z_{34}|^2}\ln\frac{\epsilon^2}{|z_{34}|^2}\,.
    \end{equation}
    Now we need to compute the $z_4$ integral
    \begin{equation}
\begin{split}
        & -2\pi\int  \frac{d^2 z_4}{z_{24}z_{34}\bz_{34}^2}\left(\ln\frac{\epsilon^2}{R^2}-\ln\frac{|z_{34}|^2}{R^2}\right)\\
        &= \frac{2\pi}{z_{23}} \int d^2 z_4 \left(\frac{1}{z_{24}\bz_{34}^2}-\frac{1}{z_{34}\bz_{34}^2}\right)\left(\ln\frac{\epsilon^2}{R^2}-\ln\frac{|z_{34}|^2}{R^2}\right)\\
        &= \frac{2\pi}{z_{23}} \int d^2 z_4 \frac{1}{z_{24}\bz_{34}^2}\left(\ln\frac{\epsilon^2}{R^2}-\ln\frac{|z_{34}|^2}{R^2}\right)\\
        &= \frac{2\pi^2}{z_{23}\bz_{23}}\left(-\ln\frac{\epsilon^2}{|z_{23}|^2}+1\right)\,.
      \end{split}
\end{equation}
    Finally we need to do the $z_3$ integral which is the same form as the one in the first integral, see \eqref{JJ.3loop.first}. Therefore the second integral equals to
    \begin{equation}
          \int  \frac{d^2 z_{3,4,5}}{z_{12}z_{13}z_{24}z_{35}z_{45}\bz_{34}\bz_{35}\bz_{45}}=\frac{\pi^3}{z_{12}^2}\left(\ln^2\frac{\epsilon^2}{|z_{12}|^2}-2\ln\frac{\epsilon^2}{|z_{12}|^2}\right)\,.
    \end{equation}

    \item The third integral is
    \begin{equation}
      \int \frac{d^2 z_{3,4,5} }{2z_{13}z_{14}z_{23}z_{25}z_{45}\bz_{34}\bz_{35}\bz_{45}}\,.
    \end{equation}
    We first integrate over $z_5$ and have
    \begin{equation}
        \int  \frac{d^2 z_5}{z_{25}z_{45}\bz_{35}\bz_{45}} = -\frac{\pi}{z_{24}\bz_{34}}\left( \ln\frac{|z_{23}|^2}{R^2} + \ln\frac{\epsilon^2}{R^2} - \ln\frac{|z_{24}|^2}{R^2} - \ln\frac{|z_{34}|^2}{R^2}\right)\,.
    \end{equation}
    Now we need to compute the $z_4$ integral
    \begin{equation}
\begin{split}
        &-\pi \int  \frac{d^2 z_4}{z_{14}z_{24}\bz_{34}^2}\left( \ln\frac{|z_{23}|^2}{R^2} + \ln\frac{\epsilon^2}{R^2} - \ln\frac{|z_{24}|^2}{R^2} - \ln\frac{|z_{34}|^2}{R^2}\right)\\
        &= \frac{\pi}{z_{12}} \int d^2 z_4 \left(\frac{1}{z_{14}\bz_{34}^2} - \frac{1}{z_{24}\bz_{34}^2}\right)\left( \ln\frac{|z_{23}|^2}{R^2} + \ln\frac{\epsilon^2}{R^2} - \ln\frac{|z_{24}|^2}{R^2} - \ln\frac{|z_{34}|^2}{R^2}\right)\\
        &= \frac{\pi^2}{z_{12}}\left[\frac{1}{\bz_{13}}\left(-\ln\frac{\epsilon^2}{|z_{12}|^2}+1+\ln\frac{|z_{13}|^2}{R^2}-\ln\frac{|z_{23}|^2}{R^2}\right)-(1\leftrightarrow2)\right]\,.
        \end{split}
\end{equation}
    Finally we need to do the $z_3$ integral and find the third integral equals to
    \begin{eqnarray}
        &&\int  \frac{d^2 z_{3,4,5} }{2z_{13}z_{14}z_{23}z_{25}z_{45}\bz_{34}\bz_{35}\bz_{45}}\nonumber\\
        &=& \pi^2\int \frac{d^2 z_4}{2z_{12}z_{13}z_{23}\bz_{13}}\left(-\ln\frac{\epsilon^2}{|z_{12}|^2}+1+\ln\frac{|z_{13}|^2}{R^2}-\ln\frac{|z_{23}|^2}{R^2}\right) + (1\leftrightarrow2)\nonumber\\
        &=& -\frac{\pi^2}{2z_{12}^2}\int d^2 z_4 \left(\frac{1}{z_{13}\bz_{13}}-\frac{1}{z_{23}\bz_{13}}\right)\left(-\ln\frac{\epsilon^2}{|z_{12}|^2}+1+\ln\frac{|z_{13}|^2}{R^2}-\ln\frac{|z_{23}|^2}{R^2}\right)+ (1\leftrightarrow2)\nonumber\\
        &=& \frac{\pi^3}{2z_{12}^2}\left(-\ln^2\frac{\epsilon^2}{|z_{12}|^2}+\ln\frac{\epsilon^2}{|z_{12}|^2} + \frac{1}{2}\ln^2\frac{\epsilon^2}{R^2} - \ln\frac{\epsilon^2}{R^2} +\frac{1}{2}\ln^2\frac{|z_{12}|^2}{R^2}\right)+ (1\leftrightarrow2)\nonumber\\
        &=& \frac{\pi^3}{z_{12}^2}\left(-\frac{1}{2}\ln^2\frac{\epsilon^2}{|z_{12}|^2}+\ln\frac{\epsilon^2}{|z_{12}|^2}\right)\,.
    \end{eqnarray}
\end{itemize}
Hence the third integral equals
\begin{equation}
 \int \frac{d^2 z_{3,4,5} }{2z_{13}z_{14}z_{23}z_{25}z_{45}\bz_{34}\bz_{35}\bz_{45}}=
 \frac{\pi^3}{z_{12}^2}\left(-\frac{1}{2}\ln^2\frac{\epsilon^2}{|z_{12}|^2}+\ln\frac{\epsilon^2}{|z_{12}|^2}\right)\,.
\end{equation}

\section{JJbar 3-loop}
\label{appendixE}
In this section we will compute in detail the four integrals in equation \eqref{JJbar3loopint}.
\begin{itemize}
    \item The first integral is
    \begin{equation}
    \label{JJbar.3loop.first}
        \int  \frac{d^2 z_{3,4,5} }{z_{13}z_{14}z_{35}z_{45}\bz_{23}\bz_{24}\bz_{35}\bz_{45}}\,.
    \end{equation}
    We first integrate over $z_5$ and have
    \begin{equation}
    \int  \frac{d^2 z_5}{z_{35}z_{45}\bz_{35}\bz_{45}} = -\frac{2\pi}{|z_{34}|^2}\ln\frac{\epsilon^2}{|z_{34}|^2}\,.
    \end{equation}
    Now we need to compute the $z_4$ integral
    \begin{equation}
    \label{JJbar.3loop.first1}
        -2\pi\ln\frac{\epsilon^2}{R^2}\int  \frac{d^2 z_4}{z_{14}z_{34}\bz_{24}\bz_{34}} + 2\pi\int  \frac{d^2 z_4}{z_{14}z_{34}\bz_{24}\bz_{34}}\ln\frac{|z_{34}|^2}{R^2}\,.
    \end{equation}
    The first term can be worked out
    \begin{equation}
     -2\pi\ln\frac{\epsilon^2}{R^2}\int  \frac{d^2 z_4}{z_{14}z_{34}\bz_{24}\bz_{34}}=
        \frac{2\pi^2}{z_{13}\bz_{23}}\ln\frac{\epsilon^2}{R^2}\left(\ln\frac{|z_{12}|^2}{R^2}+\ln\frac{\epsilon^2}{R^2}-\ln\frac{|z_{13}|^2}{R^2}-\ln\frac{|z_{23}|^2}{R^2}\right)\,.
    \end{equation}
    The second term is
        \begin{equation}
\begin{split}
        &2\pi\int  \frac{d^2 z_4}{z_{14}z_{34}\bz_{24}\bz_{34}}\ln\frac{|z_{34}|^2}{R^2}\\
        &= \frac{2\pi}{z_{13}\bz_{23}} \int d^2 z_4 \left(\frac{1}{z_{14}\bz_{24}}-\frac{1}{z_{14}\bz_{34}}-\frac{1}{z_{34}\bz_{24}}+\frac{1}{z_{34}\bz_{34}}\right)\ln\frac{|z_{34}|^2}{R^2}\\
        &= \frac{\pi^2}{z_{13}\bz_{23}} \left[\ln^2\frac{|z_{13}|^2}{R^2}+\ln^2\frac{|z_{23}|^2}{R^2}-\ln^2\frac{\epsilon^2}{R^2} + \frac{2}{\pi}I(z_1,z_2,z_3)\right]\,, 
          \end{split}
\end{equation}
    where
    \begin{equation}
        I(z_1,z_2,z_3) = \int \frac{d^2 z_4}{z_{14}\bz_{24}}\ln\frac{|z_{34}|^2}{R^2}\,,
    \end{equation}
    was introduced in \eqref{Integral123}. So, we find the $z_4$ integral gives
    \begin{eqnarray}
        &&\frac{\pi^2}{z_{13}\bz_{23}} \left[-\ln^2\frac{\epsilon^2}{|z_{12}|^2} + \ln^2\frac{\epsilon^2}{|z_{13}|^2}+\ln^2\frac{\epsilon^2}{|z_{23}|^2} + \ln^2\frac{|z_{12}|^2}{R^2} + \frac{2}{\pi} I (z_1,z_2,z_3)\right]\,.
    \end{eqnarray}
    Finally we need to do the $z_3$ integral
    \begin{equation}
         \label{JJbar.First.z3}
   \pi^2     \int  \frac{d^2 z_3}{z_{13}^2\bz_{23}^2} \left[-\ln^2\frac{\epsilon^2}{|z_{12}|^2} + \ln^2\frac{|z_{12}|^2}{R^2} + \ln^2\frac{\epsilon^2}{|z_{13}|^2}+\ln^2\frac{\epsilon^2}{|z_{23}|^2} + \frac{2}{\pi} I (z_1,z_2,z_3)\right]\,.
    \end{equation}
    All the relevant integrals needed have been computed in appendix \ref{appendix:A} and we find the first integral equal to
    \begin{equation}
         \label{JJbar.First.Final}
         \begin{split}
       & \int  \frac{d^2 z_{3,4,5} }{z_{13}z_{14}z_{35}z_{45}\bz_{23}\bz_{24}\bz_{35}\bz_{45}}\\
       & = \frac{4\pi^3}{|z_{12}|^2}\left(2-\ln\frac{\epsilon^2}{|z_{12}|^2}\right) + \pi^4\delta^{(2)}(z_{12})\left(6-4\ln\frac{\epsilon^2}{|z_{12}|^2} + \ln^2\frac{\epsilon^2}{|z_{12}|^2}\right)\,.
       \end{split}
    \end{equation}

    \item The second integral is
    \begin{equation}
    \label{JJbar.3loop.second}
        \int  \frac{d^2 z_{3,4,5} }{2z_{13}z_{15}z_{34}z_{45}\bz_{23}\bz_{24}\bz_{35}\bz_{45}}\,.
    \end{equation}
    We do the $z_5$ integral first
    \begin{equation}
        \int  \frac{d^2 z_5}{z_{15}z_{45}\bz_{35}\bz_{45}} = -\frac{\pi}{z_{14}\bz_{34}}\left(\ln\frac{|z_{13}|^2}{R^2}+\ln\frac{\epsilon^2}{R^2}-\ln\frac{|z_{14}|^2}{R^2}-\ln\frac{|z_{34}|^2}{R^2}\right)\,.
    \end{equation}
    Now we need to compute the following integral
    \begin{equation}
        -\pi\int  \frac{d^2 z_{3,4}}{2z_{13}z_{14}z_{34}\bz_{23}\bz_{24}\bz_{34}}\left(\ln\frac{|z_{13}|^2}{R^2}+\ln\frac{\epsilon^2}{R^2}-\ln\frac{|z_{14}|^2}{R^2}-\ln\frac{|z_{34}|^2}{R^2}\right)\,.
    \end{equation}
 Let us analyze the four terms in the latter expression. It is easy to see that the second together with the fourth term gives $\nicefrac14$ of the first integral, see \eqref{JJbar.3loop.first1}. We will now show that the total contribution from the first and the third term is equal to zero. We need to study
    \begin{equation}
         -\pi\int  \frac{d^2 z_{3,4}}{2z_{13}z_{14}z_{34}\bz_{23}\bz_{24}\bz_{34}}\left(\ln\frac{|z_{13}|^2}{R^2}-\ln\frac{|z_{14}|^2}{R^2}\right)\,.
    \end{equation}
    We start by doing the $z_4$ integral and find
    \begin{eqnarray}
        &&-\pi\int  \frac{d^2 z_4}{z_{14}z_{34}\bz_{24}\bz_{34}}\left(\ln\frac{|z_{13}|^2}{R^2}-\ln\frac{|z_{14}|^2}{R^2}\right)\nonumber\\
        &=& \frac{\pi^2}{z_{13}\bz_{23}}\ln\frac{|z_{13}|^2}{R^2}\left(\ln\frac{|z_{12}|^2}{R^2}+\ln\frac{\epsilon^2}{R^2}-\ln\frac{|z_{13}|^2}{R^2}-\ln\frac{|z_{23}|^2}{R^2}\right)\nonumber\\
        && + \frac{\pi}{z_{13}\bz_{23}}\int d^2 z_4 \left(\frac{1}{z_{14}\bz_{24}}-\frac{1}{z_{14}\bz_{34}}-\frac{1}{z_{34}\bz_{24}}+\frac{1}{z_{34}\bz_{34}}\right)\ln\frac{|z_{14}|^2}{R^2}\nonumber\\
        &=& \frac{\pi^2}{z_{13}\bz_{23}}\ln\frac{|z_{13}|^2}{R^2}\left(\ln\frac{|z_{12}|^2}{R^2}+\ln\frac{\epsilon^2}{R^2}-\ln\frac{|z_{13}|^2}{R^2}-\ln\frac{|z_{23}|^2}{R^2}\right)\nonumber\\
        && +\frac{\pi^2}{2z_{13}\bz_{23}}\left(-\ln^2\frac{|z_{12}|^2}{R^2}+2\ln^2\frac{|z_{13}|^2}{R^2} - \frac{2}{\pi}I(z_3,z_2,z_1)-2\ln\frac{\epsilon^2}{R^2}\ln\frac{|z_{13}|^2}{R^2}\right)\nonumber\\
        &=& \frac{\pi^2}{2z_{13}\bz_{23}}\left(2\ln\frac{|z_{12}|^2}{R^2}\ln\frac{|z_{13}|^2}{R^2}-2\ln\frac{|z_{13}|^2}{R^2}\ln\frac{|z_{23}|^2}{R^2}-\ln^2\frac{|z_{12}|^2}{R^2}-\frac{2}{\pi}I(z_3,z_2,z_1)\right)\nonumber\\
        &=& \frac{\pi^2}{z_{13}\bz_{23}}\left(\ln\frac{|z_{12}|^2}{R^2}\ln\frac{|z_{13}|^2}{R^2}-\ln\frac{|z_{13}|^2}{R^2}\ln\frac{|z_{23}|^2}{R^2} - \ln^2\frac{|z_{12}|^2}{R^2} +\ln\frac{|z_{12}|^2}{R^2}\ln\frac{|z_{23}|^2}{R^2}\right)\nonumber\\
        && +\frac{\pi^2}{2z_{13}\bz_{23}}\left(\ln^2\frac{|z_{12}|^2}{R^2}+\frac{2}{\pi}I(z_1,z_2,z_3)\right)\,.
    \end{eqnarray}
    We now multiply this by $\frac{1}{2z_{13}\bz_{23}}$ and perform the $z_3$ integral. Using the relevant integrals computed in appendix \ref{appendix:A}, we indeed find
    \begin{equation}
         -\pi\int  \frac{d^2 z_{3,4}}{2z_{13}z_{14}z_{34}\bz_{23}\bz_{24}\bz_{34}}\left(\ln\frac{|z_{13}|^2}{R^2}-\ln\frac{|z_{14}|^2}{R^2}\right) = 0\,.
    \end{equation}
So to sum up, the second integral \eqref{JJbar.3loop.second} is equal to $\nicefrac14$ of the first integral \eqref{JJbar.First.Final}
\begin{equation}
\label{JJbar.3loop.secondF}
\begin{split}
 &\int  \frac{d^2 z_{3,4,5} }{2z_{13}z_{15}z_{34}z_{45}\bz_{23}\bz_{24}\bz_{35}\bz_{45}}\\ 
 & = \frac{\pi^3}{|z_{12}|^2}\left(2-\ln\frac{\epsilon^2}{|z_{12}|^2}\right) + \frac{\pi^4}{4}\delta^{(2)}(z_{12})\left(6-4\ln\frac{\epsilon^2}{|z_{12}|^2} + \ln^2\frac{\epsilon^2}{|z_{12}|^2}\right)\,.
       \end{split}
    \end{equation}

    \item The third integral is
    \begin{equation}
        \int  \frac{d^2 z_{3,4,5} }{2z_{13}z_{14}z_{35}z_{45}\bz_{23}\bz_{25}\bz_{34}\bz_{45}}\,.
    \end{equation}
    It can be acquired from the second integral by interchanging $z_1$ and $z_2$ and then take complex conjugate and thus is equal to the second integral \eqref{JJbar.3loop.secondF}.

    \item The fourth integral is
    \begin{equation}
        \int  \frac{d^2 z_{3,4,5} }{z_{13}z_{15}z_{34}z_{45}\bz_{23}\bz_{25}\bz_{34}\bz_{45}}\,.
    \end{equation}
    We do the $z_5$ integral first and have
    \begin{equation}
        \int \frac{d^2 z_5 }{z_{15}z_{45}\bz_{25}\bz_{45}} = -\frac{\pi}{z_{14}\bz_{24}}\left(\ln\frac{|z_{12}|^2}{R^2}+\ln\frac{\epsilon^2}{R^2}-\ln\frac{|z_{14}|^2}{R^2} - \ln\frac{|z_{24}|^2}{R^2}\right)\,.
    \end{equation}
    We now need to compute the $z_4$ integral
    \begin{eqnarray}
    \label{Afterz4}
        &&-\pi\int\frac{d^2z_4}{z_{14}z_{34}\bz_{24}\bz_{34}} \left(\ln\frac{|z_{12}|^2}{R^2}+\ln\frac{\epsilon^2}{R^2}-\ln\frac{|z_{14}|^2}{R^2} - \ln\frac{|z_{24}|^2}{R^2}\right)\nonumber\\
         &=& \frac{\pi^2}{z_{13}\bz_{23}}\left(\ln\frac{|z_{12}|^2}{R^2}+\ln\frac{\epsilon^2}{R^2}-\ln\frac{|z_{13}|^2}{R^2} - \ln\frac{|z_{23}|^2}{R^2}\right)
         \left(\ln\frac{|z_{12}|^2}{R^2}+\ln\frac{\epsilon^2}{R^2}\right)\nonumber\\
        && + \frac{\pi}{z_{13}\bz_{23}}\int d^2 z_4 \left(\frac{1}{z_{14}\bz_{24}}-\frac{1}{z_{14}\bz_{34}}-\frac{1}{z_{34}\bz_{24}}+\frac{1}{z_{34}\bz_{34}}\right)\left(\ln\frac{|z_{14}|^2}{R^2} + \ln\frac{|z_{24}|^2}{R^2}\right)\nonumber\\
        &=& \frac{\pi^2}{z_{13}\bz_{23}}\left\{\left(\ln\frac{|z_{12}|^2}{R^2}+\ln\frac{\epsilon^2}{R^2}\right)\left(\ln\frac{|z_{12}|^2}{R^2}+\ln\frac{\epsilon^2}{R^2}-\ln\frac{|z_{13}|^2}{R^2} - \ln\frac{|z_{23}|^2}{R^2}\right)\right.\nonumber\\
        &&  -\ln^2\frac{|z_{12}|^2}{R^2} +\ln^2\frac{|z_{13}|^2}{R^2} -\frac{1}{\pi}I(z_1,z_3,z_2) -\frac{1}{\pi}I(z_3,z_2,z_1) + \ln^2\frac{|z_{23}|^2}{R^2}\nonumber\\
        && \left.-\ln\frac{\epsilon^2}{R^2}\ln\frac{|z_{13}|^2}{R^2}-\ln\frac{\epsilon^2}{R^2}\ln\frac{|z_{23}|^2}{R^2} \right\}\nonumber\\
        &=& \frac{\pi^2}{z_{13}\bz_{23}}\left(-\ln^2\frac{\epsilon^2}{|z_{12}|^2} + \ln^2\frac{\epsilon^2}{|z_{13}|^2} + \ln^2\frac{\epsilon^2}{|z_{23}|^2} + \ln^2\frac{|z_{12}|^2}{R^2}\right.\nonumber\\
        &&\left. -\ln\frac{|z_{12}|^2}{R^2}\ln\frac{|z_{13}|^2}{R^2} - \ln\frac{|z_{12}|^2}{R^2}\ln\frac{|z_{23}|^2}{R^2}-\frac{1}{\pi}I(z_3,z_2,z_1)-\frac{1}{\pi}I(z_1,z_3,z_2) \right).
    \end{eqnarray}
Upon using \eqref{Integral123.prop} and \eqref{Integral123.prop1} we can show that the last four terms inside the parenthesis equal to
    \begin{equation}
        \frac{2}{\pi} I(z_1,z_2,z_3)\,.
    \end{equation}
So, we are left with the integral over $z_3$
     \begin{equation}
\pi^2 \int \frac{d^2z_3}{z_{13}^2\bz_{23}^2}\left(-\ln^2\frac{\epsilon^2}{|z_{12}|^2} + \ln^2\frac{\epsilon^2}{|z_{13}|^2} + \ln^2\frac{\epsilon^2}{|z_{23}|^2} + \ln^2\frac{|z_{12}|^2}{R^2}
     + \frac{2}{\pi} I(z_1,z_2,z_3)\right)
     \end{equation}
This matches expression \eqref{JJbar.First.z3} found in the first integral \eqref{JJbar.3loop.first}; consequently, the fourth integral evaluates to \eqref{JJbar.First.Final}
    \begin{equation}
         \begin{split}
       & \int  \frac{d^2 z_{3,4,5} }{z_{13}z_{15}z_{34}z_{45}\bz_{23}\bz_{25}\bz_{34}\bz_{45}}\\
       & = \frac{4\pi^3}{|z_{12}|^2}\left(2-\ln\frac{\epsilon^2}{|z_{12}|^2}\right) + \pi^4\delta^{(2)}(z_{12})\left(6-4\ln\frac{\epsilon^2}{|z_{12}|^2} + \ln^2\frac{\epsilon^2}{|z_{12}|^2}\right)\,.
       \end{split}
    \end{equation}         
\end{itemize}

\end{document}